\documentclass[letterpaper]{article}
\usepackage{jheppub_mod}
\pdfoutput=1

\usepackage{amsmath}
\usepackage{amsfonts}
\usepackage{amsthm}
\usepackage{amssymb}
\usepackage{latexsym, array,multirow,  verbatim, enumerate}
\usepackage{slashed}
\usepackage{cancel}
\usepackage{dcolumn}
\usepackage{bm}
\usepackage{graphicx, caption, subcaption}  
\usepackage{url}
\usepackage[normalem]{ulem}
\usepackage{enumitem}
\usepackage{bbding}
\usepackage{tikz}
\usepackage{pifont}
\usepackage{empheq}
\usepackage[utf8]{inputenc}
\usepackage{bm}

\def\nn{\nonumber}

\newcommand{\ns}{_{\text{\tiny{NS}}}}

\newcommand{\ohm}{_{\text{\tiny{ohm}}}}
\newcommand{\hall}{_{\text{\tiny{hall}}}}
\newcommand{\dynamo}{_{\text{\tiny{dyn.}}}}

\newcommand{\ptl}{\partial}

\newcommand{\htt}{h^{TT}_{ij}}
\newcommand{\hx}{h_\times}
\newcommand{\hp}{h_+}

\newcommand{\orot}{\Omega_{\text{\tiny{NS}}}}

\newcommand{\mqm}{\mathcal{Q}}

\newcommand{\fD}{\mathfrak{D}}

\begin{document}

\title{Continuous Gravitational Waves and Magnetic Monopole Signatures from Single Neutron Stars }

\author[a]{P.V.S. Pavan Chandra,}
\author[b]{Mrunal Korwar}
\author[a]{and Arun M. Thalapillil}

\affiliation[a]{Indian Institute of Science Education and Research, Homi Bhabha road, Pashan, Pune 411008, India.}
\affiliation[b]{Department of Physics, University of Wisconsin-Madison, Madison, WI 53706, USA.}
\emailAdd{pvs.pavanchandra@students.iiserpune.ac.in}
\emailAdd{mkorwar@wisc.edu}
\emailAdd{thalapillil@iiserpune.ac.in}

\date{\today}

\abstract{
Future observations of continuous gravitational waves from single neutron stars, apart from their monumental astrophysical significance, could also shed light on  fundamental physics and exotic particle states. One such avenue is based on the fact that magnetic fields cause deformations of a neutron star, which results in a magnetic-field-induced quadrupole ellipticity. If the magnetic and rotation axes are different, this quadrupole ellipticity may generate continuous gravitational waves which may last decades, and may be observable in current or future detectors. Light, milli-magnetic monopoles, if they exist, could be pair-produced non-perturbatively in the extreme magnetic fields of neutron stars, such as magnetars. This non-perturbative production furnishes a new, direct dissipative mechanism for the neutron star magnetic fields. Through their consequent effect on the magnetic-field-induced quadrupole ellipticity, they may then potentially leave imprints in the early stage continuous gravitational wave emissions. We speculate on this possibility in the present study, by considering some of the relevant physics and taking a very simplified toy model of a magnetar as the prototypical system. Preliminary indications are that new-born millisecond magnetars could be promising candidates to look for such imprints. Deviations from conventional evolution, and comparatively abrupt features in the early stage gravitational waveforms, distinct from other astrophysical contributions, could be distinguishable signatures for these exotic monopole states.
}
\maketitle

\section{Introduction}

Recent observation of gravitational waves (GWs) by the LIGO-VIRGO collaboration\,\cite{Abbott:2016blz,TheLIGOScientific:2016qqj} have ushered in a new era of multi-messenger astronomy. Apart from its significant astrophysical\,\cite{TheLIGOScientific:2017qsa,Abbott:2018exr,TheLIGOScientific:2016htt} and cosmological\,\cite{Soares-Santos:2019irc,Abbott:2019yzh} implications, gravitational wave astronomy also has the potential to illuminate many important questions in fundamental physics\,\cite{Abbott:2017mem, Abbott:2018utx,Abbott:2018lct,Samajdar:2019ptt,Sathyaprakash:2019yqt}. A fast emerging area in this context is the endeavour to detect continuous GWs from single neutron stars. As opposed to GW signals from binary coalescence, which are short lived, the continuous gravitational waves are due to intrinsic deformations or other phenomena of the compact star itself, and may last decades or centuries. The cause for these continuous GWs may be due to various distinct phenomena--- stellar seismic activity, mode instabilities, mountains, oscillations or glitches in the angular velocity (see for instance\,\cite{Gualtieri:2010md, Haskell:2015psa, Glampedakis:2017nqy} and references therein). There has been rapid progress in this area, with many recent searches\,\cite{Abbott:2017ylp, Abbott:2017cvf, Authors:2019ztc}, and future third-generation GW detectors, such as the Einstein Telescope, expected to significantly improve the sensitivity and reach in the relevant frequency bands\,\cite{Punturo:2010zza,Hild:2010id,Sathyaprakash:2011bh,Sathyaprakash:2012jk}. All-sky surveys, looking for continuous gravitational waves, also hold great promise, with their ability to detect hitherto unknown sources\,\cite{Abbott:2017pqa,Pisarski:2019vxw,Dergachev:2019oyu,Dergachev:2019wqa}.

Magnetic fields are known to cause a star to become oblate or prolate, depending on the field configuration\,\cite{1953ApJ...118..116C,1954ApJ...119..407F}. This generates a quadrupole moment and associated quadrupole ellipticity. In cases where the rotation and magnetic axes do not coincide, this opens up the possibility of generating continuous gravitational waves\,\cite{Bonazzola:1995rb, Colaiuda:2007br,Ciolfi:2010td}. As opposed to gravitational waves from binary coalescences, these waveforms will last for much longer durations---days or years. This enables the application of a plethora of signal processing techniques in their analyses and understanding. The LIGO-VIRGO collaboration is already searching earnestly for such signals from pulsars\,\cite{Authors:2019ztc}. Future third-generation detectors are expected to increase the reach much further and into the niche frequency ranges of such signals\,\cite{Punturo:2010zza}. 

Magnetic monopoles have so far not been observed in nature. They are however a very generic prediction of many quantum field theories\,\cite{tHooft:1974kcl,Polyakov:1974ek} and may be awaiting discovery. Current bounds on magnetic monopoles come from colliders\,\cite{Kinoshita:1992wd, Abulencia:2005hb,Abbiendi:2007ab, MoEDAL:2016jlb}, terrestrial and balloon observations\,\cite{Ambrosio:2002qq, Hogan:2008sx, Detrixhe:2010xi}, considerations of galactic magnetic field attenuation\,\cite{Parker:1970xv,Turner:1982ag,Adams:1993fj}, searches in bulk matter\,\cite{Kovalik:1986zz, Jeon:1995rf}, and limits on monopole-catalysed proton decay in compact stars\,\cite{Kolb:1982si, Dimopoulos:1982cz, Freese:1983hz}. Very interesting limits have also been placed on heavy magnetic monopoles by considering their non-perturbative production in heavy ion collisions and in the extreme magnetic fields of neutron stars\,\cite{Gould:2017zwi}.

We are specifically interested in the case of milli-magnetic monopoles (MMM), with masses below $\mathcal{O}(1\,\mathrm{eV})$. They are monopoles with fractional effective magnetic charges, and which appear in many Standard Model extensions, especially those involving kinetic mixing\,\cite{Holdom:1985ag} with a gauge-singlet dark sector. There are previous works that have considered milli-magnetic monopoles\,\cite{Brummer:2009cs, Bruemmer:2009ky, Sanchez:2011mf,Hook:2017vyc}, in various contexts. Recently, it was also demonstrated that using energetic arguments from a magnetar, one may place very stringent, non-trivial bounds on the magnetic charge of such light MMMs\,\cite{Hook:2017vyc}. Similar bounds have also been placed on light milli-electrically charged particles\,\cite{Korwar:2017dio}, for which the relevant pair-production and astrophysical considerations are very different from MMMs.

If MMMs exist, they may be non-perturbatively pair-produced\,\cite{Affleck:1981bma,Affleck:1981ag}, via Schwinger pair-production, in the extreme magnetic fields of a neutron star, such as a magnetar\,\cite{1992ApJ...392L...9D,1993ApJ...408..194T}. This causes a decay of the magnetic field hitherto different from conventional mechanisms operational in a neutron star. The modified magnetic field evolution in turn may affect the time evolution of the quadrupole ellipticity, assuming the concerned neutron star crustal strains are below the breaking limit\,\cite{Lander:2014csa,Baiko:2018jax}. This opens up an avenue for probing these exotic states by their imprints on the gravitational waves emitted. A time evolution of the magnetic-field-induced quadrupole ellipticity, and its impact on gravitational wave emissions, has been considered previously, in other contexts\,\cite{Suvorov:2016hgr, deAraujo:2016wpz,deAraujo:2016ydk,deAraujo:2019xyn}. We would like to explore if MMMs could potentially leave markers in the gravitational waveforms, from single neutron stars, that are distinguishable from common astrophysical features.

In Sec.\,\ref{sec:cgwns} we briefly review the relevant theoretical underpinnings behind the generation of continuous gravitational waves, from single neutron stars, and outline how magnetic fields may generically lead to mass quadrupole moments. In Sec.\,\ref{sec:mmms} we then briefly review how MMMs may be incorporated in SM extensions, involving kinetic mixing, and also the relevant theoretical background on Schwinger pair production of MMMs. With the foundations laid, in Sec.\,\ref{sec:gwmmmspp} we then present our analyses and main results. We summarise and conclude in Sec.\,\ref{sec:summary}. There, we also highlight some of the shortcomings of the study, along with a few future directions.

\section{Gravitational waves from single neutron stars}{\label{sec:cgwns}}

\subsection{Continuous gravitational waves} {\label{subsec:nsdgw}}
Isolated neutron stars may emit GWs through various processes (Please see\,\cite{Glampedakis:2017nqy} and references therein for a comprehensive discussion). A neutron star may sustain a deformation in some cases, and if not axisymmetric with respect to its rotation axis, then emit GWs. Such sustained distortions, due to the elasticity of the neutron star crust\,\cite{Ushomirsky:2000ax,Owen:2005fn,Haskell:2006sv,JohnsonMcDaniel:2012wg}, are generically termed neutron star mountains. Neutron star mountains may be caused by thermal gradients\,\cite{Bildsten:1998ey,Ushomirsky:2000ax} or magnetic fields\,\cite{Bonazzola:1995rb, Haskell:2007bh, Colaiuda:2007br,Ciolfi:2010td}. We will be interested in the latter, in the context of MMMs, and will elaborate on this further in subsection\,\ref{subsec:mfinsqe}. Let us briefly review the theory behind the generation of continuous GWs, from single neutron stars, in this subsection.

In the transverse traceless gauge and an asymptotically Cartesian and mass centred coordinate system ($S$) (see \,\cite{Maggiore:1900zz,Buonanno:2007yg} for instance), the leading contribution to the gravitational wave amplitude is given by\,\cite{Ipser1971,RevModPhys.52.299}
\begin{equation}
    \htt = \frac{1}{r}\hat{\Lambda}_{ij;kl}(\Hat{n})\frac{2G}{c^4}\ddot{\mqm}_{kl}\left(t-\frac{r}{c}\right) \; .
    \label{eq:httlead}
\end{equation}
Here, for propagation direction $\hat{n}$ and $\hat{P}_{ij}(\hat{n}) = \delta_{ij} -  \hat{n}_i  \hat{n}_j$, one defines the transverse projection operator as $\hat{\Lambda}_{ij;kl} =  \hat{P}_{ik} \hat{P}_{jl} - \frac{1}{2} \hat{P}_{ij} \hat{P}_{kl}$. $\mqm$  is the mass quadrupole moment of the object. In the Newtonian limit, i.e. for weak gravitational fields, the mass quadrupole moment may be written explicitly in terms of the trace-free part of the moment of inertia tensor
\begin{equation}
    \mqm_{ij} \simeq -I_{ij} + \frac{1}{3}I_k^k\delta_{ij} \; .
\end{equation}
Here, the moment of inertia tensor $I_{ij}$ is defined in the usual way, in terms of the mass density $\rho(\mathbf{x})$, as $ I_{ij} = \int d^3x\,\rho(\mathbf{x})(x_kx^k\delta_{ij} - x_ix_j)$.
\begin{figure}
 \centering
\includegraphics[width=0.6\textwidth]{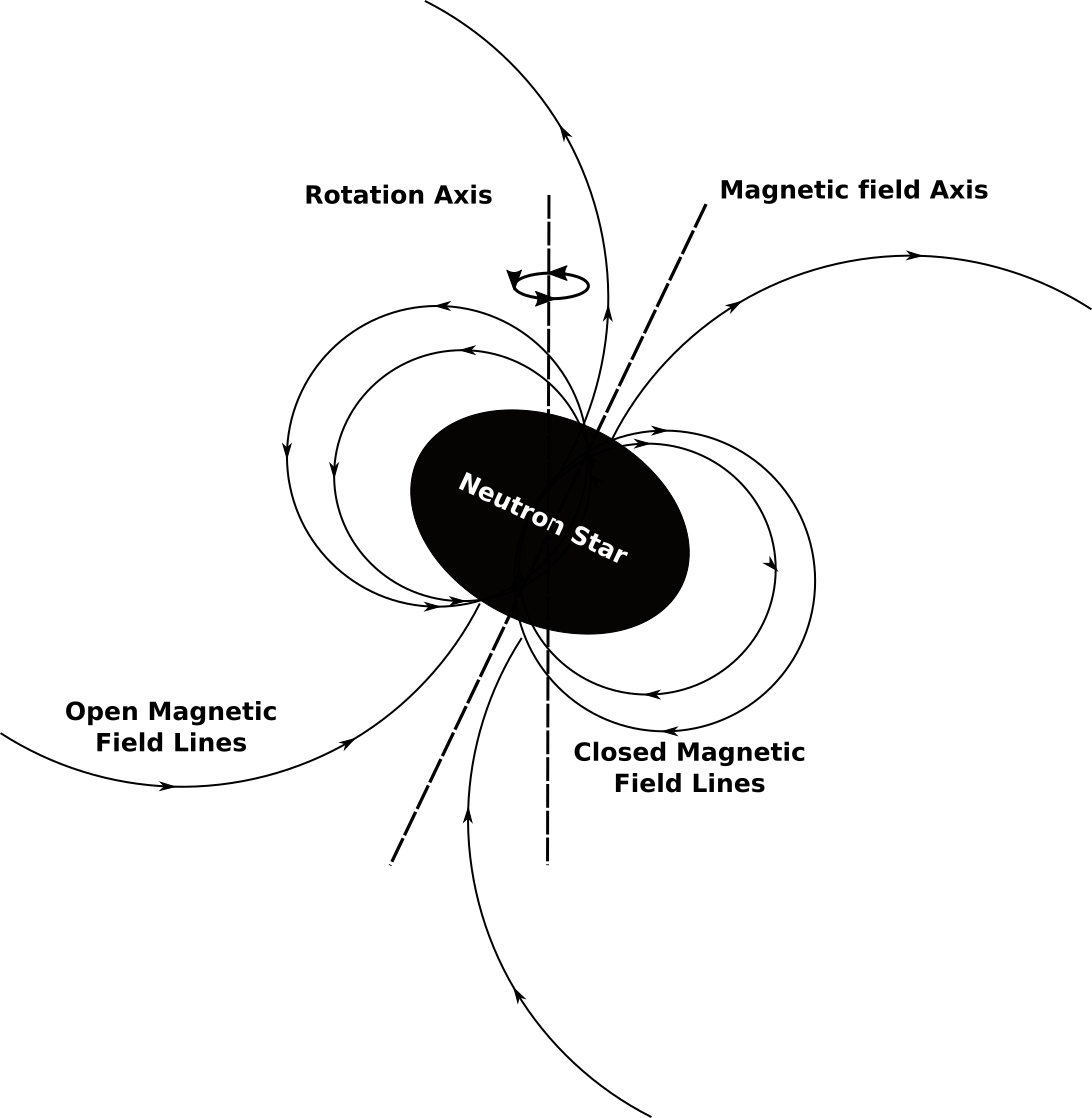}
 \caption{An illustrative representation of a neutron star, with its rotation and magnetic field axes misaligned with respect to each other. The quadrupole deformation due to the magnetic field is exaggerated for clarity. The internal field configuration is not illustrated and only the most salient features pertaining to the study are shown. The presence of a quadrupole ellipticity, with respect to the rotation axis, leads to the generation of continuous graviational waves.}
 \label{fig:ns_regions} 
 \end{figure}

Pulsars and magnetars are rotating neutron stars. If they are endowed with a quadrupole moment, there is the possibility of generating continuous GWs. The case of interest to us is where the deformations are such that there is a privileged direction---as in cases of a magnetic-field-induced deformation (see subsection\,\ref{subsec:mfinsqe}). Here, the star's magnetic moment furnishes a privileged direction, as illustrated in Fig.\,\ref{fig:ns_regions}. We also neglect any precession. Such deformations are usually parametrised either by a surface ellipticity $ \varepsilon_{\text{\tiny{S}}} = (R_{\text{\tiny{equator}}} - R_{\text{\tiny{polar}}})/R_{\text{\tiny{polar}}}$\,\cite{1953ApJ...118..116C}
or by a quadrupole ellipticity, defined as\,\cite{Bonazzola:1995rb, Colaiuda:2007br,Ciolfi:2010td}
\begin{equation}
   \boldsymbol{ \varepsilon}_{\text{\tiny{Q}}} = -\frac{\mqm}{I} \; .
    \label{eq:quadellip}
\end{equation}
Here, $I$ is the mean moment of inertia about the rotation axis, defined in terms of angular momentum $J$ as $I=J/\Omega$. In the Newtonian limit, and for a simple distortion with a privileged direction, we have the relevant $\varepsilon_{\text{\tiny{Q}}} \propto (I_{33}-I_{22})$. $\varepsilon_{\text{\tiny{S}}}$ and $\boldsymbol{\varepsilon}_{\text{\tiny{Q}}}$ quantify slightly different physics, geometrical and bulk distortions respectively, and coincide only for a star with a constant-density equation of state\,\cite{Colaiuda:2007br}. 

$\boldsymbol{\varepsilon}_{\text{\tiny{Q}}}$, which quantifies the star's  bulk deformation, is the most relevant quantity in our case. Contributions to $\boldsymbol{\varepsilon}_{\text{\tiny{Q}}}$, purely due to stellar rotations, will not contribute to continuous GWs. For the case of magnetic deformations, with the privileged direction for the deformations making two of the mass quadrupole moment eigenvalues equal, we may write the relevant quadrupole ellipticity $\tilde{\varepsilon}_{\text{\tiny{Q}}}$ as\,\cite{Bonazzola:1995rb}
\begin{equation}
\tilde{ \varepsilon}_{\text{\tiny{Q}}}  = -\frac{3}{2}\frac{\Tilde{\mqm}_{33}}{I_3} \; .
 \label{eq:qedef}
\end{equation}
Here, $\Tilde{\mqm}$ is the mass quadrupole moment due to the magnetic field, in a frame of reference ($\tilde{S}$) where it is diagonal. $I_3$ is the principal moment of inertia about the rotation axis. The $S$ and  $\tilde{S}$ coordinate system quantities are related by $\mqm = R\Tilde{\mqm}R^T$, where $R$ is an appropriate rotation matrix. The additional factor of $3/2$ is introduced to recover the classical definition of ellipticity in the Newtonian limit\,\cite{Shapiro:1983du,Bonazzola:1995rb}.

Consider now a neutron star, rotating with an angular speed $\orot$, whose rotational and magnetic field axes are misaligned by a wobble angle $\alpha$. Then, from Eq.\,(\ref{eq:httlead}), we may derive the leading GW waveform to be\,\cite{Bonazzola:1995rb}
\begin{eqnarray}
    \hp &=& h_0\sin{\alpha}\Big[\frac{1}{2}\cos{\alpha}\sin{\theta}\cos{\theta}\cos{\orot t_r}- \sin{\alpha}\frac{1 + \cos^2{\theta}}{2}\cos{2\orot t_r}\Big] \;, \nonumber \\
    \hx &=& h_0\sin{\alpha}\Big[\frac{1}{2}\cos{\alpha}\sin{\theta}\sin{\orot t_r} - \sin{\alpha}\cos{\theta}\sin{2\orot t_r}\Big] \; .
    \label{eq:hexps}
\end{eqnarray}
In the above expressions, we have defined 
\begin{equation}
h_0 = -\frac{6G}{c^4}\Tilde{\mqm}_{33}\frac{\orot^2}{r} \; .
\label{eq:hoexpq}
\end{equation}
$+$ and $\times$ denote the two polarizations. $r$ is the distance to the source and the retarded time is defined as $t_r = t - \frac{r}{c}$. $\theta$ is the line-of-sight angle to the observer, measured from the rotation axis. Through Eq.\,(\ref{eq:qedef}), note that Eq.\,(\ref{eq:hexps}) indeed has a dependence on $\tilde{\varepsilon}_{\text{\tiny{Q}}}$. From above, we see that for a general wobble angle, GWs may be emitted at $\orot$ or $2\orot$ frequencies. Eq.\,(\ref{eq:hexps}) is valid under the assumption that the magnetic field and angular velocity do not change significantly during a single period of the neutron star's rotation. This ``slow-roll" assumption is generally true for most neutron stars and will specifically be valid for the cases we study. 

The GW amplitude ($h_0$) may be directly related to the strain ($\Delta L/ L$) of the GW detector arms. The reach in $h_0$, for Advanced LIGO\footnote{\url{https://dcc.ligo.org/cgi-bin/DocDB/ShowDocument?.submit=Identifier&docid=T1800044&version=5}}  and the proposed Einstein telescope\footnote{\url{https://workarea.et-gw.eu/et/WG4-Astrophysics/base-sensitivity/et_b_spectrum.png/view}}, are around $10^{-24}-10^{-26}$ and $10^{-26}-10^{-27}$ respectively\,\cite{Gualtieri:2010md, Hild:2010id,Glampedakis:2017nqy,Authors:2019ztc}, in the $10-100\,\mathrm{Hz}$ frequency range of interest. This is assuming a year of phase-coherent observations and signal integration times\,\cite{Gualtieri:2010md, Glampedakis:2017nqy}. There have been many pioneering searches already for continuous GWs\,\cite{Abbott:2017ylp, Abbott:2017cvf, Authors:2019ztc}, and future third-generation GW detectors are expected to significantly improve the sensitivities in the niche frequency bands\,\cite{Punturo:2010zza,Hild:2010id,Sathyaprakash:2011bh,Sathyaprakash:2012jk}.

Eq.\,(\ref{eq:hexps}) may now be used in detail, to understand how the magnetic-field-induced deformations affect continuous GWs, and how specifically modifications induced by the production of MMMs will impact it. As we will remark later, we will specifically concentrate on the $2\orot$ frequency mode, without much loss of generality, for making our estimates. This choice will help us express the GW amplitude $h_0$ almost solely in terms of observable parameters, like the neutron star time period and spin-down rate.

\subsection{Magnetic field induced quadrupole moments} {\label{subsec:mfinsqe}}
Let us now briefly consider the rudimentary ideas behind stellar deformations induced by magnetic fields. It has long been known that a magnetic field threading a star could have a significant effect on its equilibrium configuration, and analogous to rotations, may induce mass quadrupole moments\,\cite{1953ApJ...118..116C,1954ApJ...119..407F}. The basic underlying physics behind this phenomena may be understood based on simple energetic arguments.

To sharpen the discussion, consider a special case for the potential deformation, in a simple model for the neutron star---a perfect sphere, of radius $R$, comprising an incompressible fluid\,\cite{1953ApJ...118..116C}. Assume that there is a uniform magnetic field in the interior and a dipolar magnetic field in the exterior. The respective field profiles are
\begin{eqnarray}
    B_r &=& B_0\cos{\theta} \quad \quad \quad \quad \quad B_{\theta} = -B_0\sin{\theta}~~~~~~~(r<R)~~~\; ,\nn\\
     B_r &=& B_0\Big(\frac{R}{r}\Big)^3\cos{\theta} \quad \quad B_{\theta} = \frac{1}{2}B_0\Big(\frac{R}{r}\Big)^3\sin{\theta} ~~~~(r>R)~~~ \; .
     \label{eq:eExtdipole}
\end{eqnarray}
Consider now a small deformation of the neutron star, parametrised as 
\begin{equation}
    r(\cos{\theta}) = R + \zeta P_l(\cos{\theta})\quad(\zeta \ll R) \; .
    \label{eq:ePldef}
\end{equation}
$P_l(\cos{\theta})$ are the Legendre polynomials. Note also in passing that $\zeta$ may be related to the surface ellipticity, through $ \varepsilon_{\text{\tiny{S}}} \sim -\zeta/R$.

If the net change in energy due to this deformation is negative, then the deformation is more stable, relative to the initial, perfectly spherical configuration. It may be shown that the non-trivial change is mainly for the spherical harmonic mode $l=2$\,\cite{1953ApJ...118..116C,1954ApJ...119..407F}, and hence we focus on this. Such quadrupole deformations are also the ones most relevant to continuous GWs.

The net change in the energy stored in the magnetic fields may be readily computed, by summing the interior and exterior contributions. This gives\,\cite{1953ApJ...118..116C}
\begin{equation}
    \delta E_{\text{\tiny{B}}} = \frac{9}{20}\zeta B_0^2R^2 \; .
    \label{etotalchangeE}
\end{equation}
Note that this is first order in $\zeta$. This change in magnetic field energy is positive if  $\zeta>0$ (prolate) and negative if $\zeta<0$ (oblate). The corresponding change in gravitational energy, due to the deformation, is
\begin{equation}
    \delta E_{\text{\tiny{G}}}  = \frac{3}{25}\left(\frac{\zeta}{R}\right)^2\frac{GM^2}{R}\; .
    \label{egravchange}
\end{equation}
Note that in contrast to $ \delta E_{\text{\tiny{B}}}$, this is second order in $\zeta$ and is thus always positive. The total change in energy is obtained by summing the magnetic and gravitational energy contributions. This gives
\begin{equation}
    \delta E = \frac{3}{25}\left(\frac{\zeta}{R}\right)^2\frac{GM^2}{R} + \frac{9}{20}\zeta B_0^2R^2
\end{equation}
Note from above that, for $\zeta \ll R$, the sign of the net change in energy will be determined directly by the sign of $\zeta$. 

To obtain the most stable configuration, we need to minimise $\delta E$, and if it comes out to be negative, would suggest an energetically more favorable configuration\,\cite{1953ApJ...118..116C}. Minimisation gives
\begin{equation}
    \frac{\bar{\zeta}}{R} = -\frac{15}{8}\frac{B_0^2R^4}{GM^2}=-\frac{9}{2}\Big(\frac{B_0}{B_*}\Big)^2\; .
\end{equation}
Here, $B_*^2 = 12 GM^2/ 5R^4$ is the limit on the magnetic field coming from the virial theorem\,\cite{1953ApJ...118..116C}, and corresponds to around $10^{18}\,\mathrm{G}$ for neutron stars. Thus, under this magnetic configuration, the incompressible fluid star undergoes an oblate deformation, departing from pure spherical symmetry. This is the basic idea behind how quadrupole moments are generated by magnetic fields threading a star. This is in fact a generic phenomena, with the exact nature and extent of the deformation depending on the magnetic field configuration and the star's specific equation of state.

 For an external dipolar magnetic field configuration in a neutron star, let us now examine a few simple equation of states, and their effects on bulk deformation (quantified by $\tilde{\varepsilon}_{\text{\tiny{Q}}}$). To simplify discussions, define a dimensionless deformation parameter ($\fD$) through the relation
\begin{equation}
 \tilde{\varepsilon}_{\text{\tiny{Q}}}  = \fD\frac{B^2}{B_*^2} \;.
 \label{eq:DeformDef}
\end{equation}
Without loss of generality, we have made the normalisation with respect to $B_*$. The deformation parameter $\fD$, may be related to the magnetic distortion factor defined in\,\cite{Bonazzola:1995rb}. 

Consider the case of a constant density fluid. In this case, the quadrupole ellipticity may be computed as\,\cite{Haskell:2007bh}
\begin{equation}
     \tilde{\varepsilon}_{\text{\tiny{Q}}}^{\text{\tiny{const.}}}  = \frac{2}{15}\frac{B^2}{B_*^2}\;,
  \end{equation}
  giving $\fD=2/15$. For the case of an $n=1$ polytrope, again with an exterior dipolar magnetic field, we have\,\cite{Haskell:2007bh}.
\begin{equation}
\tilde{\varepsilon}_{\text{\tiny{Q}}}^{\text{\tiny{1-poly.}}}    = \frac{36\pi^5(12-\pi^2)}{5(\pi^2 - 6)^3}\frac{B^2}{B_*^2}\; ,
\end{equation}
in which case $\fD=\frac{36\pi^5(12-\pi^2)}{5(\pi^2 - 6)^3}$. For almost the same magnetic field magnitude and exterior field configuration, the latter polytropic equation of state leads to a larger deformation.

 Considering the values of the deformation parameter, in these examples, it seems $\fD\sim [10^{-1},10^{2}]$. These ranges for  $\fD$ are also believed to be typical for more realistic equation of states and field configurations\,\cite{Bonazzola:1995rb,Haskell:2007bh}, and we will use them for making our estimates. The effects due to rotations have been neglected in these estimates\,\cite{Haskell:2007bh}.
 
 There are a few observational upper bounds on $ \tilde{\varepsilon}_{\text{\tiny{Q}}}$, for neutron stars in their early stages. X-ray light curves from short gamma ray bursts have been used to constrain $ \tilde{\varepsilon}_{\text{\tiny{Q}}}$ of post-merger stable neutron stars, giving mean bounds in the range\,\cite{Lasky:2015olc,Glampedakis:2017nqy}
 
 \begin{equation}
 \tilde{\varepsilon}_{\text{\tiny{Q}}}^{_{\text{\tiny{Obs. GRB}}}}  \lesssim 10^{-2}-10^{-1} \; .
 \label{eq:qelimitgrb}
 \end{equation}
 
 For pulsars in their later stages, there are constraints from continuous GW searches by the LIGO-VIRGO collaboration, giving fiducial ellipticity bounds in the range $[10^{-2},10^{-8}]$\,\cite{Abbott:2017ylp, Abbott:2017cvf, Authors:2019ztc}. Theoretical models suggest bounds on fiducial ellipticities of compact stars in the range $10^{-2}-10^{-7}$\,\cite{Ushomirsky:2000ax,Owen:2005fn,Haskell:2006sv,Ciolfi:2010td,JohnsonMcDaniel:2012wg}; depending on the stellar mass, hadron composition, epoch, equation of state and theoretical approximations used. Interestingly, there is even possibly an indication for a lower bound on $\tilde{\varepsilon}_{\text{\tiny{Q}}} $, of about $10^{-9}$, from analyses of millisecond pulsars\,\cite{Woan:2018tey}. We will always work with values well below the mean bounds in Eq.\,(\ref{eq:qelimitgrb}). The main difference from taking lower values for $\tilde{\varepsilon}_{\text{\tiny{Q}}}$, or equivalently $\fD$, will be to make the GW signal undetectable much earlier in time, since the neutron star's birth; or completely undetectable if $\fD$ is exteremely small.
 
 In summary, the elastic properties of the neutron star crust\,\cite{Ushomirsky:2000ax,Owen:2005fn,Haskell:2006sv,Ciolfi:2010td,JohnsonMcDaniel:2012wg}, and presence of very strong magnetic fields, may lead generically to the presence of sustained deformations, resulting in a non-zero quadrupole ellipticity. As remarked earlier, there may even be a time evolution of the magnetic-field-induced quadrupole ellipticity in these early phases. This is a plausible scenario assuming that the concerned crustal stresses and strains, due to the magnetic pressure, are below the breaking limit\,\cite{Lander:2014csa,Baiko:2018jax}. An evolving quadrupole ellipticity has been previously studied, in other GW contexts\,\cite{Suvorov:2016hgr, deAraujo:2016wpz,deAraujo:2016ydk,deAraujo:2019xyn}, and we would like to explore if the presence of MMMs may leave imprints on this quadrupole ellipticity evolution, and consequent GW generation.

\section{ Milli-magnetic monopoles and non-perturbative production}{\label{sec:mmms}

\subsection{Milli-magnetic monopoles and theoretical foundations} {\label{subsec:mmm}}

Magnetic monopoles are yet to be observed in nature. They nevertheless seem to be a very generic prediction of many quantum field theories and model frameworks (see for instance \,\cite{Preskill:1984gd}, and related references). 

In conventional Maxwellian electrodynamics, the homogeneous equation $\vec{\nabla}\cdot\vec{B} = 0$, or equivalently the Bianchi identity of the field tensor $F_{\alpha\beta}$, presupposes the non-existence of magnetic monopoles. In this framework, the manifestly covariant equations in vacuum take the form
\begin{equation}
    \ptl_\mu F^{\mu\nu} = 0~,~~\ptl_\mu\Tilde{F}^{\mu\nu} = 0\;.
\end{equation}
Here, $\Tilde{F}^{\mu\nu} = \frac{1}{2}\epsilon^{\mu\nu\rho\sigma}F_{\rho\sigma}$ is the dual field tensor, and the Bianchi identity implies $F^{\mu\nu} = \ptl^{\mu}A^\nu - \ptl^\nu A^\mu$. As is well know, the vacuum equations are symmetric under the duality transformation
\begin{eqnarray}
    F^{\mu\nu}\rightarrow \Tilde{F}^{\mu\nu}~,~~\Tilde{F}^{\mu\nu}\rightarrow -F^{\mu\nu}\;.
\end{eqnarray}

Once we introduce an electric source, say $J^\alpha$, this symmetry is lost. To consider restoration of the symmetry, we may speculate the addition of an analogous magnetic source term $K^\alpha$. The equations then take the form
\begin{equation}\label{eq:maxwelleq}
    \ptl_\mu F^{\mu\nu} = -eJ^\nu~,~~\ptl_\mu\Tilde{F}^{\mu\nu} = -g K^\nu \;,
\end{equation}
which are clearly symmetric under the transformations
\begin{eqnarray}
    F^{\mu\nu}\rightarrow \Tilde{F}^{\mu\nu}~&,&~~\Tilde{F}^{\mu\nu}\rightarrow -F^{\mu\nu}\nn\\
    e J^\nu\rightarrow g K^\nu~&,&~~ g K^\nu\rightarrow - eJ^\nu \; .
\end{eqnarray}
The addition of the $K^\alpha$ term introduces magnetic monopoles.

The theoretical underpinnings for milli-magnetic monopoles, in the context of kinetic mixings, were discussed in\,\cite{Hook:2017vyc}, and put on a firmer theoretical foundation later in\,\cite{Terning:2018lsv}. Among the theoretical subtleties, in incorporating magnetic monopoles directly in a quantum field theory, is the fact that it is not possible to write a local, Lorentz invariant Lagrangian containing both electric and magnetic charges\,\cite{DIRAC:1948,Hagen:1965zz,Zwanziger:1970hk,PhysRevD.18.2080}. We briefly review the theoretical framework\,\cite{Terning:2018lsv} for incorporating MMMs, through kinetic mixing, as a specific example of incorporating MMMs into beyond Standard Model extensions. This will also help fix notations. 

 One theoretical strategy to incorporate magnetic monopoles, by Zwanziger\,\cite{Zwanziger:1970hk}, contains two gauge potentials $A_{\alpha}$ and $\tilde{A}_{\alpha}$, with a local Lagrangian, but without any manifest Lorentz invariance \cite{Zwanziger:1970hk,Csaki:2010rv}. In this formulation, one of the gauge potentials, $A_{\alpha}$, couples locally to the electric current $J_{\alpha}$, while the other, $\tilde{A}_{\alpha}$, couples to the magnetic current $K_{\alpha}$. The Lagrangian density takes the form \,\cite{Zwanziger:1970hk,Csaki:2010rv,Terning:2018lsv}
\begin{align}
\mathcal{L} &= -\frac{n^{\alpha}n^{\mu}}{2 n^{2}} \Big{[}\eta^{\beta\nu}\big{(}F^{A}_{\alpha\beta}F^{A}_{\mu\nu}+F^{\tilde{A}}_{\alpha\beta}F^{\tilde{A}}_{\mu\nu} \big{)} - \frac{1}{2}\epsilon_{\mu}{}^{\nu\gamma\delta}\big{(}F^{\tilde{A}}_{\alpha\nu}F^{A}_{\gamma\delta}-F^{A}_{\alpha\nu}F^{\tilde{A}}_{\gamma\delta} \big{)}\Big{]}  \nonumber   \\  &- e J_{\mu} A^{\mu} - \frac{4\pi}{e} K_{\mu}\tilde{A}^{\mu} \; .
\end{align}
Here, $F^{A}_{\alpha\beta}=\partial_{\alpha}A_{\beta}-\partial_{\beta}A_{\alpha}$ and $F^{\tilde{A}}_{\alpha\beta}=\partial_{\alpha}\tilde{A}_{\beta}-\partial_{\beta}\tilde{A}_{\alpha}$ are the respective field tensors. $n_{\alpha}$ is an arbitrary four vector, corresponding to the direction of the Dirac string in certain gauge choices. The presence of $n_{\alpha}$, projects out two, on-shell photon polarizations, breaking manifest Lorentz-invariance \cite{Zwanziger:1970hk,Csaki:2010rv, Terning:2018lsv}. It has been argued that physical observables of the theory are independent of $n_{\alpha}$\,\cite{PhysRevD.18.2080}. The above Lagrangian density correctly gives the modified Maxwell's equations in Eq.\,(\ref{eq:maxwelleq}), with the definition 
\begin{equation}
F_{\mu\nu} = \frac{n^{\alpha}}{n^{2}} \big{(}n_{\mu}F^{A}_{\alpha\nu}-n_{\nu}F^{A}_{\alpha\mu}-\varepsilon_{\mu\nu\alpha}{}^{\beta}n^{\gamma}F^{\tilde{A}}_{\gamma\beta} \big{)} \; .
\label{eq:mft}
\end{equation}

Let us now understand how MMMs may specifically be included, in this framework, in the context of kinetic mixing\,\cite{Holdom:1985ag}. For this, consider the Lagrangian density\,\cite{Terning:2018lsv} incorporating kinetic mixing with a dark sector (whose low-energy states are all Standard Model gauge singlets; labelled by subscript `$\text{D}$')
\begin{align} \label{eq:kineticMixing}
\mathcal {L}_{\text{\tiny{MMM}}} &\supset -\frac{n^{\alpha}n^{\mu}}{2 n^{2}} \Big{[}\eta^{\beta\nu}\big{(}F^{A}_{\alpha\beta}F^{A}_{\mu\nu}+F^{\tilde{A}}_{\alpha\beta}F^{\tilde{A}}_{\mu\nu} \big{)} - \frac{1}{2}\epsilon_{\mu}{}^{\nu\gamma\delta}\big{(}F^{\tilde{A}}_{\alpha\nu}F^{A}_{\gamma\delta}-F^{A}_{\alpha\nu}F^{\tilde{A}}_{\gamma\delta} \big{)}\Big{]}  - e J_{\mu} A^{\mu} - \frac{4\pi}{e} K_{\mu}\tilde{A}^{\mu} \nonumber \\ & 
-\frac{n^{\alpha}n^{\mu}}{2 n^{2}} \Big{[}\eta^{\beta\nu}\big{(}F^{A}_{\text{\tiny{D}}\alpha\beta}F^{A}_{\text{\tiny{D}}\mu\nu}+F^{\tilde{A}}_{\text{\tiny{D}}\alpha\beta}F^{\tilde{A}}_{\text{\tiny{D}}\mu\nu} \big{)} - \frac{1}{2}\epsilon_{\mu}{}^{\nu\gamma\delta}\big{(}F^{\tilde{A}}_{\text{\tiny{D}}\alpha\nu}F^{A}_{\text{\tiny{D}}\gamma\delta}-F^{A}_{\text{\tiny{D}}\alpha\nu}F^{\tilde{A}}_{\text{\tiny{D}}\gamma\delta} \big{)}\Big{]}  \nonumber   \\  
& - \frac{m^{2}_{\text{\tiny{D}} A}}{2}A_{\text{\tiny{D}}\mu}A^{\mu}_\text{\tiny{D}} - e_\text{\tiny{D}} J_{\text{\tiny{D}}\mu} A_\text{\tiny{D}}^{\mu} - \frac{4\pi}{e_\text{\tiny{D}}} K_{\text{\tiny{D}}\mu}\tilde{A}_\text{\tiny{D}}^{\mu} + \chi \frac{n^{\alpha}n^{\mu}}{n^{2}}\eta^{\beta\nu}\big{(}F^{A}_{\text{\tiny{D}}\alpha\beta}F^{A}_{\mu\nu}-F^{\tilde{A}}_{\text{\tiny{D}}\alpha\beta}F^{\tilde{A}}_{\mu\nu} \big{)} \; .
\end{align}
$F^{A}_\text{\tiny{D}}$ and $F^{\tilde{A}}_\text{\tiny{D}}$ are the field tensors corresponding to the dark gauge potentials $A_\text{\tiny{D}}$ and $\tilde{A}_\text{\tiny{D}}$. $J_\text{\tiny{D}}$ and $K_\text{\tiny{D}}$ are the dark electric and magnetic currents, with $e_\text{\tiny{D}}$ being the dark electric charge. $e$ and $e_\text{\tiny{D}}$ are in general independent parameters of the model. Without loss of generality,  we take the $n_{\alpha}$ four-vector to be the same in both the sectors; this can always be achieved with appropriate gauge transformations. The two sectors are connected by kinetic mixing, via the last term in Eq.\,(\ref{eq:kineticMixing}). This term is equivalent to $\chi/2 F_{\mu\nu}F^{\mu\nu}_\text{\tiny{D}}$, from the definition in Eq.\,(\ref{eq:mft}). The mass term for $A_{\text{\tiny{D}}\mu}$ breaks the $SO(2)$ symmetry of the kinetic terms and is uniquely responsible for MMMs\,\cite{Terning:2018lsv}.

Considering $A_\text{\tiny{D}}^{\mu}$ to be massive, after field redefinitions, we get magnetic monopoles that have effective milli-magnetic charges\,\cite{Hook:2017vyc,Terning:2018lsv}, at low energies. Explicitly, consider the field redefinitions
\begin{eqnarray}
 A_{\mu}\rightarrow  A_{\mu} + \chi A_{\text{\tiny{D}}\mu} ~&,&~~\tilde{A}_{\mu}\rightarrow \tilde{A}_{\mu} \nn\\
A_{\text{\tiny{D}}\mu}\rightarrow  A_{\text{\tiny{D}}\mu} ~&,&~~ \tilde{A}_{\text{\tiny{D}}\mu}\rightarrow \tilde{A}_{\text{\tiny{D}}\mu}-\chi \tilde{A}_{\mu}  \; .
\end{eqnarray}
Note that the above field transformations, ensure that the visible-sector gauge potentials ($A_{\mu},\, \tilde{A}_{\mu}$) do not get mass terms, and hence $U(1)_{\text{EM}}$ remains unbroken. After these field redefinitions, making the kinetic terms canonical, the relevant interaction terms become
\begin{equation}
\mathcal{L}_{\text{\tiny{int.}}} \supset e J_{\mu}A^{\mu} + e \chi J_{\mu}A_\text{\tiny{D}}^{\mu} + e_\text{\tiny{D}} J_{\text{\tiny{D}}\mu}A_\text{\tiny{D}}^{\mu} +  \frac{4\pi}{e}K_{\mu}\tilde{A}^{\mu} + \frac{4\pi}{e_\text{\tiny{D}}}K_{\text{\tiny{D}}\mu}\tilde{A}_\text{\tiny{D}}^{\mu} - \frac{4\pi\chi}{e_\text{\tiny{D}}}K_{\text{\tiny{D}}\mu}\tilde{A}^{\mu} \;.
\label{eq:effint}
\end{equation}

After making the kinetic terms canonical, one now has an effective interaction of the form $4\pi\chi/e_\text{\tiny{D}} \,K_{\text{\tiny{D}}\mu}\tilde{A}^{\mu}$. This makes the dark-sector magnetic monopoles milli-magnetically charged under the visible photon, with an interaction strength of $4\pi\chi/e_\text{\tiny{D}}$. $\chi$ in general is an arbitrary, irrational number. This is the origin of the fractional magnetic charge, and of MMMs. Naively, $\chi$ being an irrational number may seem to violate the Dirac charge quantization condition at low energies. The emergence of milli-magnetically charged particles, through kinetic mixing, is nevertheless still consistent with a global Dirac quantization condition\,\cite{Brummer:2009cs,Terning:2018lsv}.

Moving forward, let us henceforth define all MMM charges with respect to the visible sector $g \equiv 4 \pi/e$. Towards this end, define the MMM charge parameter $\xi$ as
\begin{equation}
\xi \equiv \,\chi \left(\frac{g_\text{\tiny{D}}}{g}\right) \;.
\label{eq:mmmcharge}
\end{equation}
Here, we have defined $g_\text{\tiny{D}}\equiv4 \pi/e_\text{\tiny{D}}$. With respect to our photon, MMMs therefore have magnetic charges $\xi g \equiv \chi g_\text{\tiny{D}} $. We will express all analyses and limits with respect to $\xi$ henceforth.

\subsection{Non-perturbative pair production of milli-magnetic monopoles} {\label{subsec:sppmmm}}

In Quantum Electrodynamics, when the field strengths are very large, one may have non-perturbative production of electrically or magneticallly charged particles, through the Schwinger pair-production mechanism\,\cite{Sauter:1931zz,Heisenberg1936,Schwinger:1951nm,Affleck:1981bma,Affleck:1981ag}. This is a distinct phenomena compared to, for instance, perturbative electron-positron pair-production ($\gamma+\gamma \rightarrow e^+ + e^-$). For field strengths comparable to the particle masses, the non-perturbative rates may be exponentially enhanced. 
\begin{figure}
    \centering
    \includegraphics[scale=0.0875]{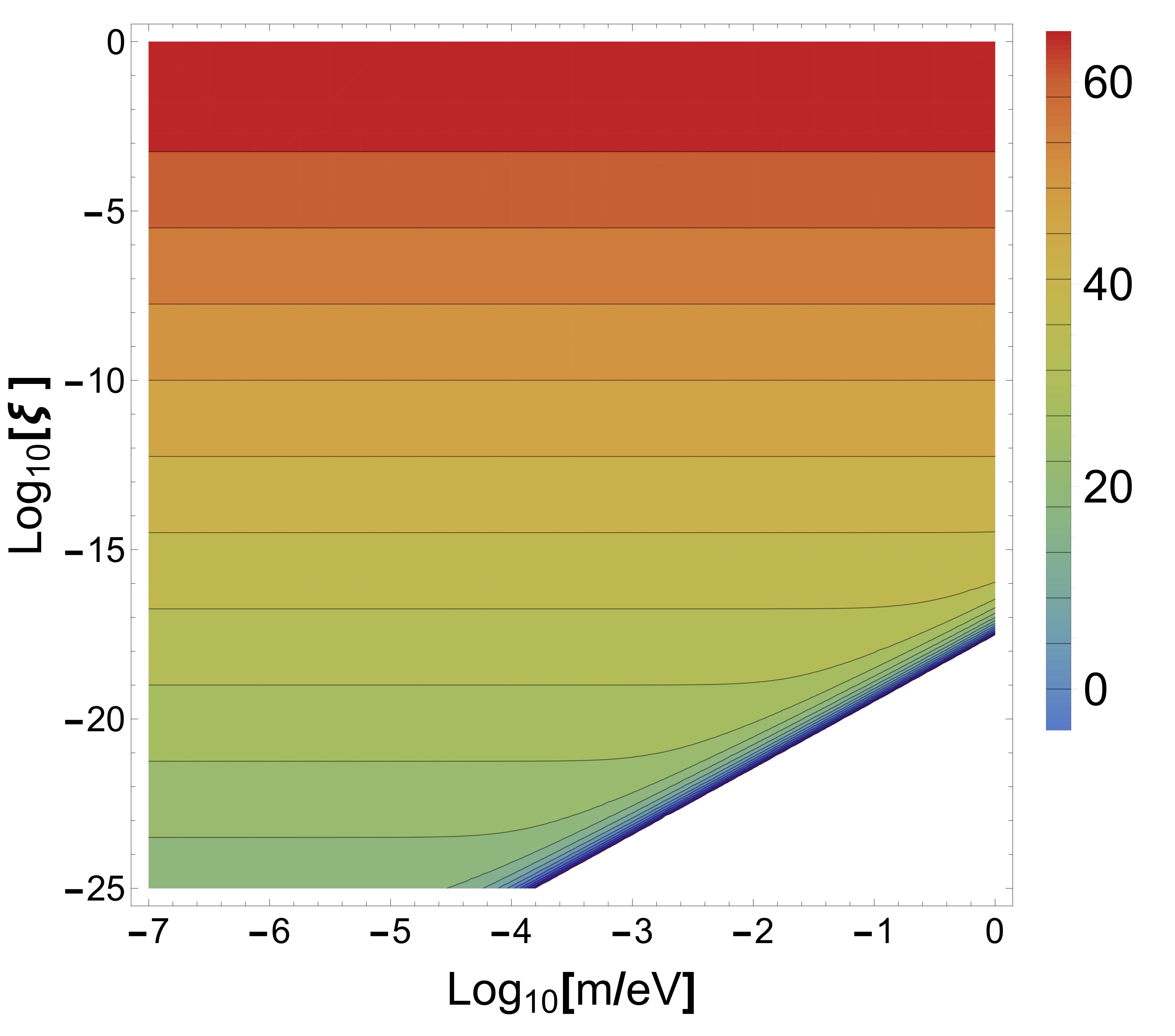}
    \caption{The pair-production rates per unit volume ($\log_{10} \left[\Gamma_{0}/1\, \text{m}^{-3} \text{s}^{-1} \right]$), for milli-magnetic monopoles at zero temperature, are shown. The magnetic field has been taken to be $10^{16}\,\mathrm{G}$. The zero temperature rates bracket the true rates that may be operational in systems with a finite temperature.}
    \label{fig:0Gamma}
\end{figure}

For zero temperature and  homogeneous magnetic fields, as compared to the Compton wavelength and separation of the particles, the average MMM pair-production rate, per unit volume, is given by\,\cite{Affleck:1981bma,Affleck:1981ag}
\begin{equation} 
\Gamma_{0}= \frac{\xi^2 g^2 B^{2}}{8\pi^{3}} \exp\Bigg{[}-\frac{\pi m^{2}}{\xi g B}\Bigg{]} \; .
\label{eq:ztmspp}
\end{equation} 
The zero temperature rate assuming a magnetic field of $10^{16}\,\mathrm{G}$ is shown in Fig.\,\ref{fig:0Gamma}. This is the first term in the vacuum decay rate\,\cite{Affleck:1981bma,Affleck:1981ag,Cohen:2008wz}. Recently, this computation was also extended to strong coupling and finite temperatures\,\cite{Gould}. 

We are interested in light, milli-magnetically charged monopoles of mass $m\ll \mathcal{O}(1\,\mathrm{eV})$, with effective magnetic charges $\xi g \ll 1$, as in Eq. (\ref{eq:effint}). We assume that $g_\text{\tiny{D}} \lesssim g \equiv 4 \pi/e$, and that any higher order instanton corrections to the MMM pair-production rates\,\cite{Affleck:1981bma,Affleck:1981ag,Cohen:2008wz,Gould} may be neglected, to good approximation. Also note that for the MMM mass ranges we consider, the Compton wavelengths ($\lambda_{\text{\tiny{Compt.}}}^{\text{\tiny{max}}}  \lesssim 1\,\mathrm{m}$) are such that local magnetic field inhomogeneities in the neutron star may be neglected, to leading order.

Based on theoretical models and measurements, currently observed neutron stars are believed to have mean surface temperatures of the order of $10^6\,\mathrm{K}$. It is believed that in the early stages of their formation, the mean temperatures may have been even higher ($\sim10^{11}\,\mathrm{K}$). In the standard cooling scenario for neutron stars, it is presumed that a neutron star when formed has internal temperatures approaching $10^{11}\, \mathrm{K}$ or more, and subsequently cools down by various processes---neutrino emissions (through the URCA and modified URCA processes), neutrino pair bremsstrahlung, thermal photon emissions and so on (see, for instance,\,\cite{Shapiro:1983du,Becker:2009} and references therein). The rate of cooling differs widely during the many stages, with timescales  varying from seconds to thousands of years. The neutron star mean temperature is thought to evolve from around $10^{11}\, \mathrm{K}$ to $10^{4}\, \mathrm{K}$ over a few million years\,\cite{Shapiro:1983du,Becker:2009}.

Thus, a more relevant quantification of the MMM production rate, at least in the initial phases of the neutron star's life, should try to incorporate the effects of this finite temperature. As mentioned earlier, there has been tremendous progress recently in computing Schwinger pair-production rates at finite temperature, both for electrically charged as well as for strongly-coupled magnetic monopoles\,\cite{Dittrich:1979ux, Elmfors:1993wj,Gies:1998vt, Gies:1999vb,Ganguly:1998ys,Kim:2010qq, Brown:2015kgj, Medina,Gould:2017zwi,Gould,Gould:2018ovk,Korwar:2018euc,Draper:2018lyw}. There is currently some disagreement on the exact functional form of the worldline instanton (see for instance discussions in\,\cite{Brown:2015kgj, Medina,Gould:2017zwi,Gould,Gould:2018ovk,Korwar:2018euc,Draper:2018lyw}). Nevertheless, there seem to be a few generic predictions---an exponential enhancement in the pair-production rate relative to zero temperature rates, and a critical temperature below which the thermal enhancements switch off\,\cite{Gies:1998vt,Gies:1999vb,Brown:2015kgj,Medina,Gould:2017zwi,Gould,Korwar:2018euc}. 

The critical temperature ($T_{\text{\tiny{C}}}$) is a function of the magnetic field, monopole mass and magnetic charge\,\cite{Gies:1998vt,Gies:1999vb,Brown:2015kgj,Medina,Gould:2017zwi,Gould,Korwar:2018euc}
\begin{equation}
T_{\text{\tiny{C}}} (m,\xi,B) \equiv \frac{\xi g B}{2m} \; .
\label{eq:ct}
\end{equation}
Below this critical temperature, the thermal enhancements turn off and the rate subsequently follows the zero temperature rate, given by Eq.\,(\ref{eq:ztmspp}). The critical temperature estimates for our regions of interest are illustrated in Fig.\,\ref{fig:CritTemp}.

The thermal rate, at a finite temperature $T\equiv\beta^{-1}$, may be approximated as~\cite{Medina, Korwar:2018euc}
\begin{eqnarray}  
&\Gamma_{\text{\tiny{T}}}&\big(m,\xi,B,T\big) \simeq  \sum_{p=1}^{\infty} \frac{(-1)^{p+1}\xi^2 g^{2}B^{2}}{8\pi^{3}p^{2}} \exp\Bigg{[}-\frac{p\pi m^{2}}{\xi g B}\Bigg{]}  + \Theta(T-T_{\text{\tiny{C}}}) \sum_{p=0}^{\infty}\sum_{n=1}^{n_{max}} 2(-1)^{p} \frac{ (\xi g B)^{2}}{(2\pi)^{3/2}(nm\beta)^{1/2}\vartheta^{2}} \nn \\
	&& \Big{[}1-\Big{(}\frac{n\beta \xi g B}{2m}\Big{)}^{2}\Big{]}^{-\frac{1}{4}} \exp\Bigg{[}-\frac{m^{2}}{2 \xi g B}\Big{[}2\pi (p+1) - 2\arcsin\big{(}\frac{nT_{\text{\tiny{C}}}}{T}\big{)}\Big{]} +\frac{nm}{2T}\sqrt{1-\frac{n^{2}T_{\text{\tiny{C}}}^{2}}{T^{2}}}\Bigg{]}  \; ,
\label{eq:tgamma}
\end{eqnarray} 
following the notion of an electromagnetic dual to Schwinger pair production by an electric field\,\cite{Affleck:1981bma,Gould,Medina, Korwar:2018euc}. Here, $\Theta(x)$ is the Heaviside step function, $n_{max}\equiv \lfloor2 m/(\xi g B\beta)\rfloor=\lfloor T/T_{\text{\tiny{C}}}\rfloor$, and $\vartheta = 2\pi(p+1)-2\arcsin\Big{(}\frac{nT_{\text{\tiny{C}}}}{T}\Big{)}$. $\lfloor x \rfloor$ denotes the integer less than or equal to $x$. This explicit analytic expression derived in the worldline instanton framework, utilising a saddle-point approximation, is valid for the semi-classical parameter $\xi g B/m^2 \lesssim 2 \pi$~\cite{Affleck:1981bma, Dunne:2005sx, Medina, Gould, Korwar:2018euc}. Note that the enhancement is present only when $T> T_{\text{\tiny{C}}}$, as already mentioned, and changes abruptly below it. In fact, Eq.\,(\ref{eq:tgamma}) seems to suggest that the rate also changes abruptly at all integer multiples of $T_{\text{\tiny{C}}}$, owing to $n_{max}=\lfloor T/T_{\text{\tiny{C}}}\rfloor$. We will utilise the above expression, in regions satisfying $\xi g B/m^2 \lesssim 2 \pi$, to estimate Schwinger pair-production rates at finite temperatures. 

Note that at a characteristic worldline sphaleron temperature, much higher than $T_{\text{\tiny{C}}}$, the pair production transitions from a quantum tunnelling phenomena to a classical, thermal process, described by a worldline sphaleron\,\cite{Gould:2018ovk}. The characteristic worldline sphaleron temperature\,\cite{Gould:2018ovk}, where this transition occurs, is greater than $\sim 10^{11}\,\mathrm{K}$ for the parameter space of interest to us. Since the neutron star is believed to cool to around $10^{11}\,\mathrm{K}$ within just a few seconds of its formation, we are mostly outside the sphaleron regime. 
\begin{figure}
    \centering
    \includegraphics[scale=0.0875]{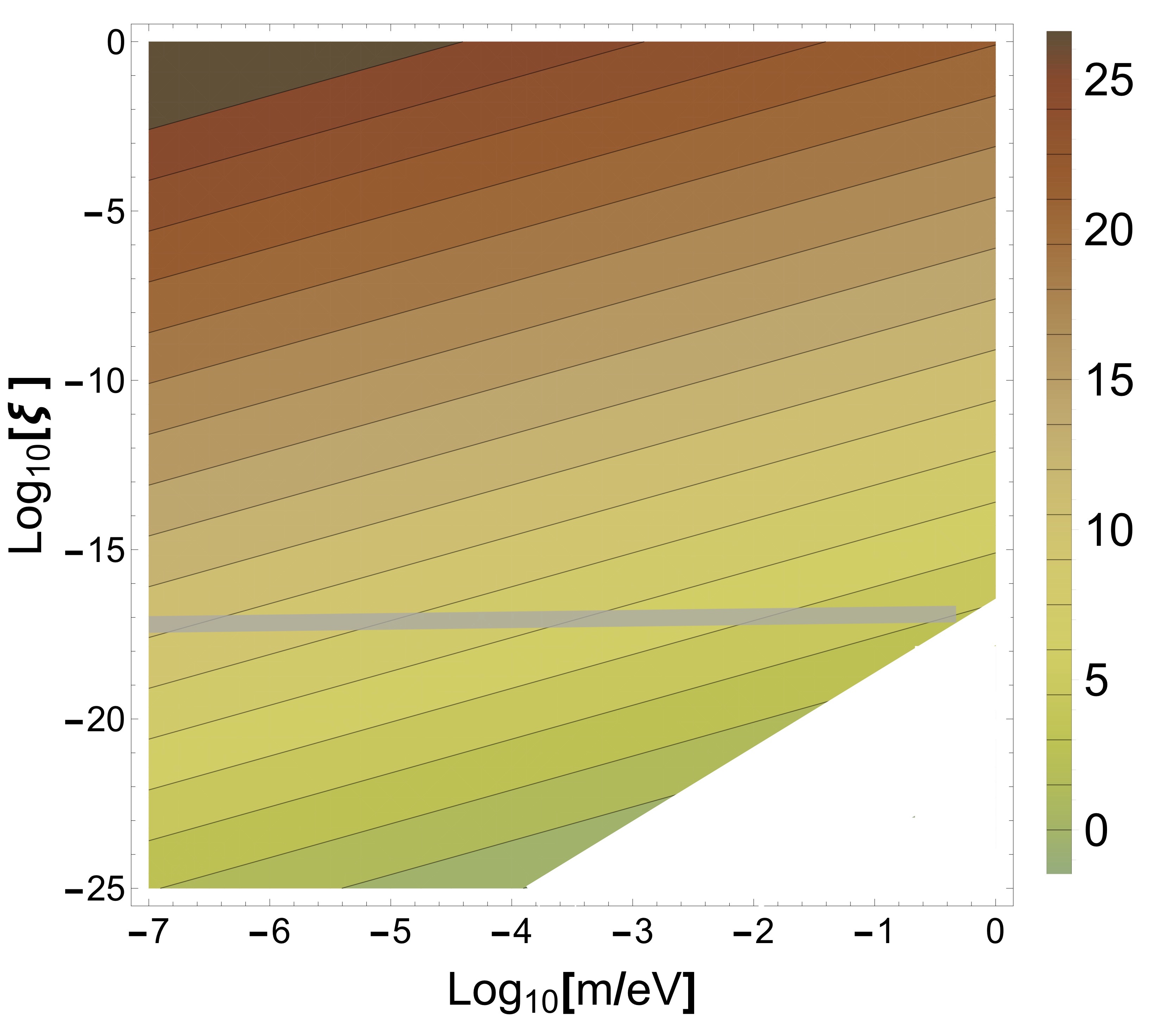}
    \caption{Plot of $\log_{10}\left[T_{\text{\tiny{C}}}/1\,\mathrm{K}\right]$ is shown, for a fixed magnetic field of $10^{16}\,\mathrm{G}$. Certain regions are irrelevant, due to the exponential suppression of Schwinger pair-production rates. Mean energetic arguments from magnetars\,\cite{Hook:2017vyc}, also render regions with $\xi \gtrsim 10^{-17}$ (gray band) unviable,  for $m\lesssim 1\,\mathrm{eV}$.}
    \label{fig:CritTemp}
\end{figure}

For the MMM and dark photon mass ranges we will consider, the MMM Compton wavelength and string separation between monopole and anti-monopole\,\cite{Shifman:2002yi,Hook:2017vyc,Terning:2018lsv} are also such that the magnetic field spatial inhomogeneities may be neglected, to good approximation. The temporal variation of the magnetic field is also very gradual, and its effects may similarly be neglected while computing rates, to leading order. 

The additional magnetic field dissipation, due to Schwinger pair production of MMMs, may cause a deviation in the  time evolution of the gravitational wave amplitude, and frequency, relative to the conventional case. The fact that the non-perturbative pair-production rate reverts to the zero temperature rate, below a characteristic temperature $T_{\text{\tiny{C}}}$\,\cite{Gies:1998vt,Gies:1999vb,Brown:2015kgj,Medina,Gould:2017zwi,Gould,Korwar:2018euc}, also opens up an intriguing possibility. As the neutron star cools down during its lifetime, if milli-magnetic monopoles exist, there could potentially be an abrupt change in the monopole production rate, in the vicinity of $T_{\text{\tiny{C}}}$, that relatively brusquely affects the gravitational wave amplitude and frequency subsequent to it. As emphasised before, $T_{\text{\tiny{C}}}$ itself is a function of the magnetic field, monopole mass and magnetic charge $\xi$. Note that as the MMMs we are considering have very small masses and tiny magnetic charges, we do not expect them to drastically affect the ordinary thermal evolution or dynamic processes in the neutron star in a very significant way.

These comparatively abrupt features in the waveform would be a universal signature, potentially visible across different magnetar systems, in their early phase continuous gravitational wave emissions. They should also be distinct from signals originating due to typical astrophysical phenomena, and hence potentially distinguishable. As may be deduced from Fig.\,\ref{fig:CritTemp}, for a field of $10^{16}\,\mathrm{G}$, the critical temperature may be as high as $10^{8}\,\mathrm{K}$, in the viable $(m,\xi)$ parameter space of interest.

\section{Effects of milli-magnetic monopoles on gravitational waves}{\label{sec:gwmmmspp}}
With the basic concepts in place from the previous sections, we may now undertake a study of what potential affects MMMs may have on continuous gravitational waves from single neutron stars.

The MMMs are generally confined objects with a string connecting the monopole and anti-monopole\,\cite{Shifman:2002yi,Hook:2017vyc,Terning:2018lsv}. They behave like magnetically charged objects only beyond a particular distance $\mathcal{O}(1/m_\text{\tiny{D}A})$. This suggests a characteristic lower value for the dark photon mass $m_\text{\tiny{D}A}$. There is also an upper bound to $m_\text{\tiny{D}A}$ that must be considered. The external magnetic field will accelerate the MMMs out of the magnetar, as long as the string tension between the pair produced MMMs ($\mathcal{O}(m_\text{\tiny{D}A}^2)$) is smaller than the external electromagnetic force. The gravitational forces on the MMMs, due to the neutron star, are many orders of magnitude smaller than the Lorentz forces, and hence do not furnish any further bounds. These requirements altogether translate finally to\,\cite{Hook:2017vyc}
\begin{equation}
 \frac{1}{R\ns} \lesssim m_\text{\tiny{D}A} \lesssim  \sqrt{\xi g B}    \; .
\end{equation}

For the parameter space of interest, the upper bound gives $m_\text{\tiny{D}A} \lesssim 10^8 \,\mathrm{km}^{-1}$, which may be trivially incorporated. Neutron stars have typical radii $\sim 10\,\mathrm{km}$ and we set the lower limit for the dark photon mass by it. This will also make robust our assumption of magnetic field homogeneity, relative to the particle Compton wavelength and separation. We will work assuming the above two bounds for $m_\text{\tiny{D}A}$.  Lower dark photon masses and corresponding modifications may be readily incorporated phenomenologically, by assuming an exponential suppression\,\cite{Hook:2017vyc} of the external field, as felt by the monopole and anti-monopole. 

The subsequent history of the MMMs, after they are pair-produced and expelled by the magnetic field, is not important, as they do not return energy back into the magnetic fields. As mentioned earlier, due to the tiny MMM mass and charge, any direct imposition on the thermal or dynamical evolution of the neutron star should also be very marginal, after production. This is in sharp contrast to heavy magnetic monopoles, if they exist, that may be captured and trapped by neutron stars, and which may impact the internal neutron star processes and dynamics more drastically. For instance, these heavy magnetic monopoles may efficiently catalyse nucleon decays in the neutron star\,\cite{Kolb:1982si, Dimopoulos:1982cz, Freese:1983hz}. It is also distinct from interesting scenarios where very heavy dark matter states could be captured by neutron stars, sometimes through multiple scatterings, heating them up kinetically or through subsequent annihilations\,\cite{Bramante:2017xlb,Raj:2017wrv}. In such cases, measuring the temperatures of very old neutron stars could lead to very interesting constraints\,\cite{Bramante:2017xlb,Raj:2017wrv}.

It was pointed out recently, in\,\cite{Hook:2017vyc}, that by considering an average magnetar field of $10^{15}\,\mathrm{G}$, monopole anti-monopole pair-production rates bracketed by the zero temperature rate, and an assumed magnetar active lifetime of $10^4\, \mathrm{yrs}$, one may place strong bounds on viable MMMs. For magnetars with magnetic fields in the range $10^{15}-10^{16}\,\mathrm{G}$, and for various dark photon masses, such energetic considerations give limit estimates of
\begin{equation}
\xi \lesssim 10^{-17}\;,
\label{eq:mmmcl}
\end{equation}
for  $m \lesssim \mathcal{O}(1\,\mathrm{eV})$. Following\,\cite{Hook:2017vyc}, we will explicitly compute the limit on $\xi$ and impose it, at each MMM mass of interest, before utilising that point to study the evolution of the gravitational wave amplitude.

Let us now turn to the GW waveforms that could be expected. To be concrete, let us focus specifically on the GW mode with frequency $2\orot$. Assuming the dominance of electromagnetic dipole radiation, from Eq.(\ref{eq:hexps}), the amplitude corresponding to the $2\orot$  frequency mode may be expressed as 
\begin{eqnarray}
    h^{2\orot, +}_0 &=&\frac{8}{5}\fD\frac{R\ns^2}{cr}\frac{\Dot{P}}{P}\frac{1+\cos^2{\theta}}{2} \; ,\nn \\
    h^{2\orot, \times}_0 &=&\frac{8}{5}\fD\frac{R\ns^2}{cr}\frac{\Dot{P}}{P}\cos{\theta} \; .\
    \label{eq:hpcamps}
\end{eqnarray}
Note that when expressed in terms of the observables $\dot{P}$ and $P$ in this fashion, the amplitude at frequency $2\orot$, is independent of the moment of inertia and the unknown wobble angle $\alpha$. This is an advantage to considering this specific frequency mode, as we had alluded to earlier. There is a dependence on the line-of-sight angle $\theta$, that just gives an $\mathcal{O}(1)$ factor, and may be ignored for our order of magnitude estimates. The dominance of electromagnetic dipole radiation may be explicitly checked for reasonable values of $ \tilde{\varepsilon}_{\text{\tiny{Q}}}$, and we shall comment further on this later. 

From Eq.\,(\ref{eq:hpcamps}), the order of magnitude estimate for the GW amplitude gives
\begin{equation}
    h^{2\orot}_0 \simeq 10^{-31}\,\fD\,\Big(\frac{R\ns}{10\, \mathrm{km}}\Big)^2\, \Big(\frac{\textrm{kpc}}{r}\Big)\, \Big(\frac{\textrm{s}}{P}\Big)\, \Big(\frac{\Dot{P}}{10^{-11}}\Big) \; .
    \label{eq:hoamp}
\end{equation}
 As we had remarked earlier, in subsec.\,\ref{subsec:nsdgw}, the sensitivity in strain ($h_0$) for Advanced LIGO and the proposed Einstein telescope, are around $10^{-24}-10^{-26}$ and $10^{-26}-10^{-27}$ respectively\,\cite{Gualtieri:2010md, Hild:2010id,Glampedakis:2017nqy,Authors:2019ztc}, in the $10-100\,\mathrm{Hz}$ frequency range of relevance to these continuous GWs. This is assuming 1-year signal integration times\,\cite{Gualtieri:2010md, Glampedakis:2017nqy}. We note therefore from above that the amplitude is typically very small, except when the compact object is spinning rapidly, undergoing rapid braking with large $\dot{P}$ or has large magnetic field induced deformations. One may therefore intuit, from Eq.\,(\ref{eq:hoamp}), that one must search for candidate compact stars with aforementioned characteristics. 

This may be further sharpened by estimating the typical GW amplitudes one may expect from observed pulsars and magnetars, due to their assumed magnetic-field-induced quadrupole ellipticities, for reasonable ranges of the deformation parameter $\fD$. These estimates are shown in Fig.\,\ref{fig:mgatnfestimates}, for a few representative pulsar and magnetar candidates. The parameter values were taken from the ATNF\footnote{\url{https://www.atnf.csiro.au/research/pulsar/psrcat/}} pulsar\,\cite{Manchester:2004bp} and McGill\footnote{\url{http://www.physics.mcgill.ca/~pulsar/magnetar/main.html}} magnetar\,\cite{Olausen:2013bpa} catalogues. Estimates in Fig.\,\ref{fig:mgatnfestimates} suggest that magnetars with large time periods ($\sim 10\,\mathrm{s}$) and conventional radio pulsars with relatively small magnetic fields ($\sim 10^{11}\,\mathrm{G}$), or equivalently small $\dot{P}$, may not be the most promising candidates to look for persistent GWs; or for that matter MMM imprints in them.
\begin{figure}
    \centering
    \includegraphics[scale = 0.555]{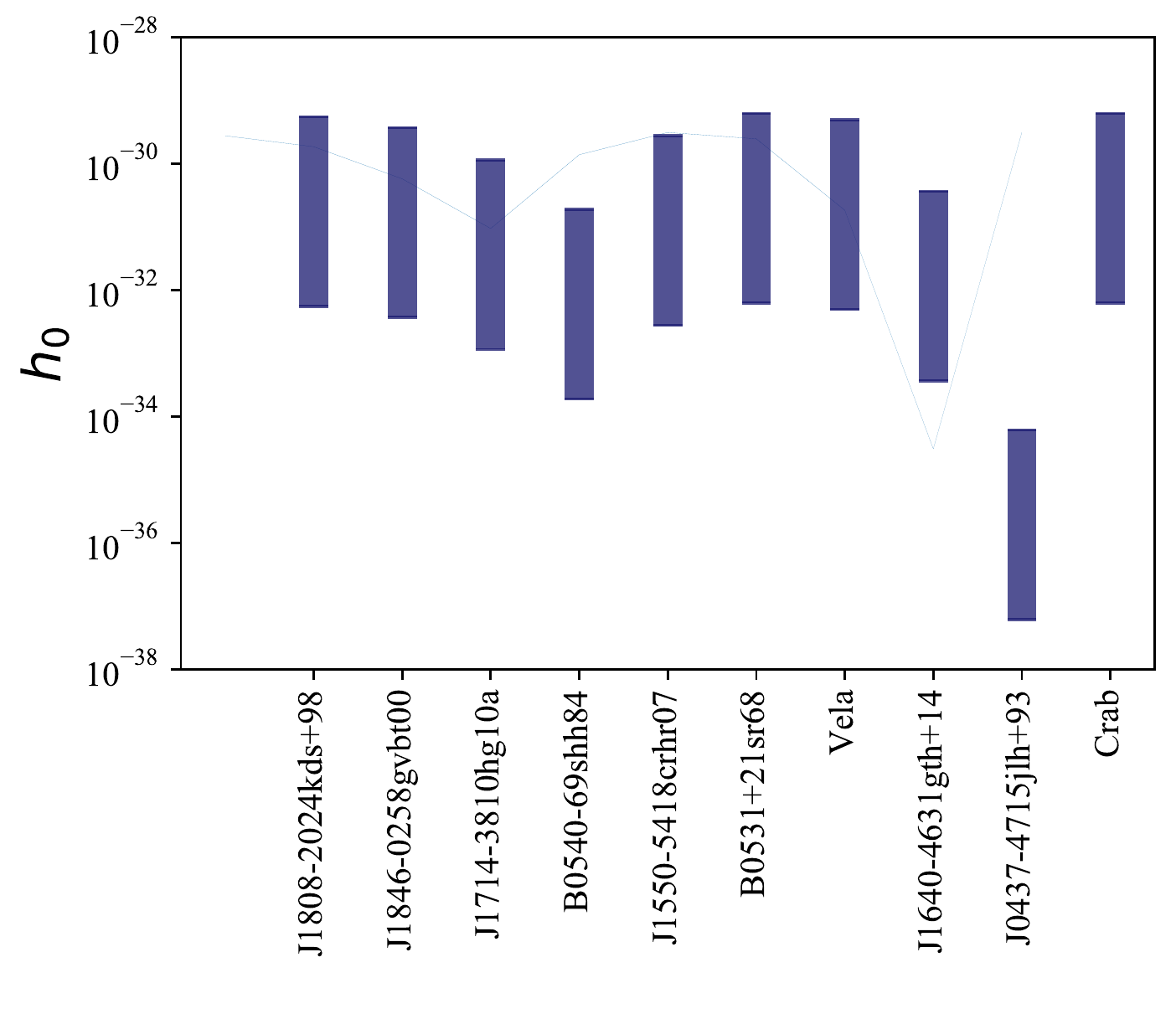}
    \includegraphics[scale = 0.525]{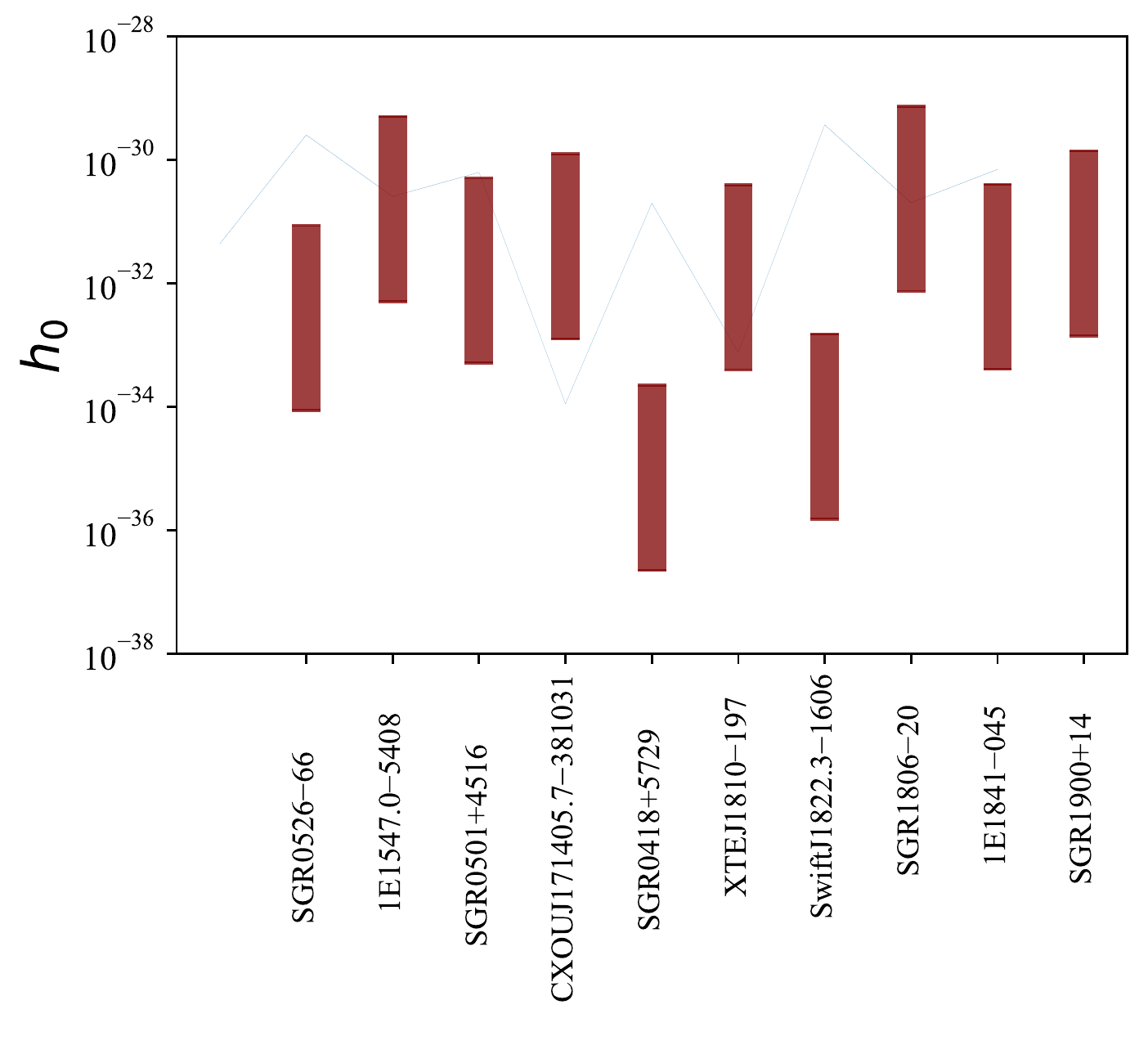}
    \caption{Estimates for the magnetic-field-induced GW amplitudes, from a few representative pulsar (Left) and magnetar (Right) candidates. The relevant parameter values were taken from the ATNF pulsar\,\cite{Manchester:2004bp} and McGill magnetar\,\cite{Olausen:2013bpa} databases. $\fD$ is varied in the range $[10^{-1},10^{2}]$.}
    \label{fig:mgatnfestimates}
\end{figure}

Based on these broad inspections, perhaps the most promising candidates are a class of newly-born magnetars, in their early stages of evolution---the so called  millisecond magnetars\,\cite{Dai:1998bb,Zhang:2000wx,Rowlinson:2013ue,Giacomazzo:2013uua,2010ApJ...717..245K,2011MNRAS.413.2031M,0004-637X-798-1-25,Metzger:2017wdz}. Millisecond magnetars are new-born neutron stars with very high magnetic fields and very small time periods, and have already been speculated to be promising sources for continuous GWs\,\cite{Gualtieri:2010md, Giacomazzo:2013uua,Glampedakis:2017nqy}. They have also garnered much interest recently, in the context of fast radio bursts\, \cite{Metzger:2017wdz,Suvorov:2019rzz}. The other reason for optimism, while considering these candidates, is that the internal magnetic fields and temperatures are presumed to be much higher, during the early stages of the magnetar's formation; relative to their mean values taken over the entire magnetar lifetime. This opens up the possibility that detectable signatures may still be present in the early stages. The mean temperature of the neutron star is also varying very rapidly in the early epochs, and as we shall discuss later, this increases the possibility of MMM induced abrupt features in the GW waveforms. We therefore explore imprints on gravitational waves from millisecond magnetars, induced by MMMs; with magnetic charges below the bound set by mean energetic limits, as in Eq.\,(\ref{eq:mmmcl}). 

Let us therefore look at the effects of MMM non-perturbative pair production in a very simplified toy model, for a  newly-born millisecond magnetar. Consider specifically the magnetic field evolution in this toy model, assuming an external dipolar and uniform internal magnetic field, that attempts to capture the salient features. The simplified evolution equation\,\cite{Gould:2017zwi,1992ApJ...395..250G,Aguilera:2007xk,Vigano:2013lea,2007A&A...470..303P,1993ApJ...408..194T} may be written as
\begin{equation}
    \frac{dB\ns(t)}{dt} \simeq \frac{B\ns(t)}{\tau\dynamo}e^{-{t}/{\tau\dynamo}}-\frac{B\ns(t)}{\tau\ohm}-\frac{B\ns^2(t)}{B\ns(0)\tau\hall} - \frac{2\xi g \,l\,V_{m}}{R\ns^3}\,\Gamma_{\text{\tiny{T}}}\big(m,\xi,B\ns(t),T(t)\big)  \; .
    \label{eq:MagFieldEvol}
\end{equation}

The various terms try to crudely encapsulate the characteristic time-scales of the various relevant processes that are operational. 

The first term is a dynamo term\,\cite{1993ApJ...408..194T}, that is believed to be operational for the first few seconds of a neutron star's birth, after which it winds down. It amplifies and regenerates the magnetic field in the magnetar. The second and third terms are the Ohmic and Hall drift terms, that contribute conventionally to the decay of the magnetic fields in a neutron star. Following standard literature, we take the dynamo, Ohmic and Hall drift time constants as $\tau\dynamo=10\,\mathrm{s}$, $\tau\ohm=10^6\,\mathrm{yrs}$ and $\tau\hall=10^4\,\mathrm{yrs}$\,\cite{1992ApJ...395..250G,Aguilera:2007xk} respectively. The respective time constants are in reality non-trivial functions of temperature and density, but the above values have been found to capture relevant effects\,\cite{Aguilera:2007xk}. A toy model of the magnetic field evolution, as encapsulated by Eq.\,(\ref{eq:MagFieldEvol}), has also been seen to semi-quantitaively reproduce\,\cite{Aguilera:2007xk} essential results from more detailed magneto-thermal simulations\,\cite{Aguilera:2007xk,Vigano:2013lea,2007A&A...470..303P}. A similar evolution equation was also considered recently in \,\cite{Gould:2017zwi}, to set interesting limits on strongly-coupled, heavy magnetic monopoles.

The last term in Eq.\,(\ref{eq:MagFieldEvol}) is due to the Schwinger pair production of MMMs, and is derived from energy conservation arguments. Specifically, it is obtained by equating the loss of energy from the electromagnetic field, to the energy needed for Schwinger pair production and to the work done in accelerating the monopole anti-monopole pairs outward.  $V_m$ is the active volume over which MMMs are being non-perturbatively pair produced, and is taken to be the volume of the neutron star. $l$ is the mean distance over which MMMs are being accelerated by the magnetic field, after production, and is equated to the diameter of the neutron star. The Schwinger pair production of the MMMs causes a non-perturbative decay of the magnetic flux. This is a potentially new source of flux decay in neutron stars, different from classical processes. Energy is being expended from the magnetic field during pair-production and during their expulsion.

Eq.\,(\ref{eq:MagFieldEvol}) must be solved in tandem with the neutron star spin-down equation
\begin{equation}
    \frac{d\orot(t)}{dt} \simeq -\frac{5}{12}\frac{R\ns^4}{M\ns} B\ns^2(t)\orot^3(t) - \frac{64}{25}G M\ns R\ns^2 \tilde{\varepsilon}_{\text{\tiny{Q}}}^2(t)\orot^5(t) \;.
    \label{eq:OmegaEvol}
\end{equation}
In this spin down equation, we have assumed that the magnetic axis is orthogonal to the rotation axis, i.e., $\alpha = \frac{\pi}{2}$\,\cite{0004-637X-798-1-25}. Note from Eq.\,(\ref{eq:hexps}) that this choice would also cause continuous gravitational emissions solely at $2\orot$ frequencies. In the above expression, the neutron star has been idealised to an almost spherical object, with moment of inertia $\sim \frac{2}{5} M\ns R\ns^2$. The first term in Eq.\,(\ref{eq:OmegaEvol}) is due to electromagnetic dipole radiation, and the second term incorporates the gravitational quadrupole radiation. The latter term incorporates braking due to GW emissions and is proportional to $\tilde{\varepsilon}_{\text{\tiny{Q}}}^2(t)$. The GW emission contribution is small compared to the dipole term, for all $ \tilde{\varepsilon}_{\text{\tiny{Q}}}$ values of interest to us, as may be explicitly verified. It hence validates the assumption in Eq.\,(\ref{eq:hpcamps}). We neglect effects due to precession, in the time evolution.

When there is non-perturbative pair production of MMMs, the full gravitational waveform is plausibly affected, relative to the conventional case, in both amplitude and frequency. As seen from Eqs.\,(\ref{eq:qedef}),\,(\ref{eq:hexps}), (\ref{eq:hoexpq}) and\,(\ref{eq:DeformDef}), the amplitude of the waveform is modified directly due to the refinement of the quadrupole ellipticity. It is also affected indirectly through the adjustments in $\orot(t)$, induced via the modified magnetic field evolution of Eq.\,(\ref{eq:MagFieldEvol}) and by the GW emission term in Eq.\,(\ref{eq:OmegaEvol}). The latter effects also modify the frequency of the emitted gravitational waveform $2\orot(t)$.
\begin{eqnarray}
h_0(t) ~&\propto&  \tilde{\varepsilon}_{\text{\tiny{Q}}}(t) \, \orot(t)^2 \;,\nn \\
\dot{\Omega}\ns(t) &\propto&  B\ns^2(t)~,~~ \tilde{\varepsilon}_{\text{\tiny{Q}}}^2(t)\; .
\end{eqnarray}
Remembering that $\tilde{\varepsilon}_{\text{\tiny{Q}}}(t) \propto B\ns(t)^2$, ultimately all the altered characteristics are a consequence of the MMM modified magnetic field evolution, condensed in the simplified Eq.\,(\ref{eq:MagFieldEvol}). Thus, a revised modulation in the frequency and amplitude envelope of the GW waveform should be a consequence of MMM production in general. 

On a related note, observe from Eq.\,(\ref{eq:MagFieldEvol}) that during the first many seconds after the millisecond magnetar's birth (say around time $t_0$) one may in some instances have a steady state situation ($\dot{B}\ns(t_0) \sim 0$). This may be prompted by a near cancellation of the positive dynamo and negative MMM contributions
\begin{equation}
\frac{B\ns(t_0)}{\tau\dynamo}e^{-{t_0}/{\tau\dynamo}} \sim  \frac{2\xi g \,l\,V_{m}}{R\ns^3}\,\Gamma_{\text{\tiny{T}}}\big(m,\xi,B\ns(t_0),T(t_0)\big)\; .
\end{equation}
This quasi steady-state, if achieved, should also reflect in the persistent GW emissions during these brief intervals; before the dynamo shuts off after $\mathcal{O}(10\,\mathrm{s})$. The time-scales for the Ohmic and Hall-drift processes are much longer, and should not play a significant role at these very early times. The possibility of such a steady state was also effectively leveraged in\,\cite{Gould:2017zwi}, to place very interesting lower bounds on the mass of heavy magnetic monopoles.

To explore further, we numerically solve Eqs.\,(\ref{eq:MagFieldEvol}) and (\ref{eq:OmegaEvol}), with a starting point taken as $10\,\mathrm{yrs}$ after the millisecond magnetar formation\,\cite{Dai:1998bb,Zhang:2000wx,Rowlinson:2013ue,Giacomazzo:2013uua,2010ApJ...717..245K,2011MNRAS.413.2031M,Metzger:2017wdz}; in a binary neutron star merger or supernovae explosion. For the estimates, initial starting values of $B\ns^0=10^{16}\,\mathrm{G}~,~~\orot^0=2 \pi/(30\,\mathrm{ms})$ and $T^0_{\text{\tiny{NS,pole}}}=4.5\times10^6\,\mathrm{K}$, as well as temperature evolution profiles, are taken following representative values in the literature\,\cite{Vigano:2013lea,2010ApJ...717..245K,2011MNRAS.413.2031M,Metzger:2017wdz}. The neutron star equatorial temperature is usually much lower than the polar temperature\,\cite{Vigano:2013lea} and the internal temperatures are believed to be much higher. Discounting magnetic fields, the interior temperature is thought to be related to the surface temperature via an approximate scaling that roughly goes as $T_{\text{\tiny{NS,in}}} \sim T^2_{\text{\tiny{NS,surf.}}}$\,\cite{1983ApJ...272..286G}. To reduce model assumptions, to the extent possible, we will take the neutron star polar temperature prediction\,\cite{Vigano:2013lea} as a crude proxy for the mean neutron star temperature. Assumption of a higher mean temperature would cause a further enhancement to the thermal Schwinger pair-production rate, and would only cause more pronounced deviations from conventional evolution. $\fD$ is taken to be $81$, corresponding to the case of an $n=1$ polytropic equation of state. This gives an initial $\tilde{\varepsilon}_{\text{\tiny{Q}}}$ of about $10^{-4}$. This magnitude seems to be consistent with typical expectations, for millisecond magnetars\,\cite{Suvorov:2019rzz}. The distance to the source is taken as $1\,\mathrm{kpc}$. For a magnetic charge of $\xi=10^{-19}$, the  MMM masses have been taken to be $15\,\mathrm{meV}$, $20\,\mathrm{meV}$, and $25\,\mathrm{meV}$. The magnetic charge adopted for these masses, satisfies the limit from mean energetic arguments, as derived in \,\cite{Hook:2017vyc}. The parameter space points also satisfy $\xi g B/m^2 \lesssim 2 \pi$, making Eq.\,(\ref{eq:tgamma}) valid, and hence directly usable in Eq.\,(\ref{eq:MagFieldEvol}). The dark photon mass has been taken as  $m_\text{\tiny{D}A}=10^3\,\mathrm{m}^{-1}$, which is consistent with current limits (See for instance discussions in\,\cite{Collar:2012olx, Jaeckel:2010ni}, and references therein).
\begin{figure}
    \centering
    \includegraphics[scale = 0.65]{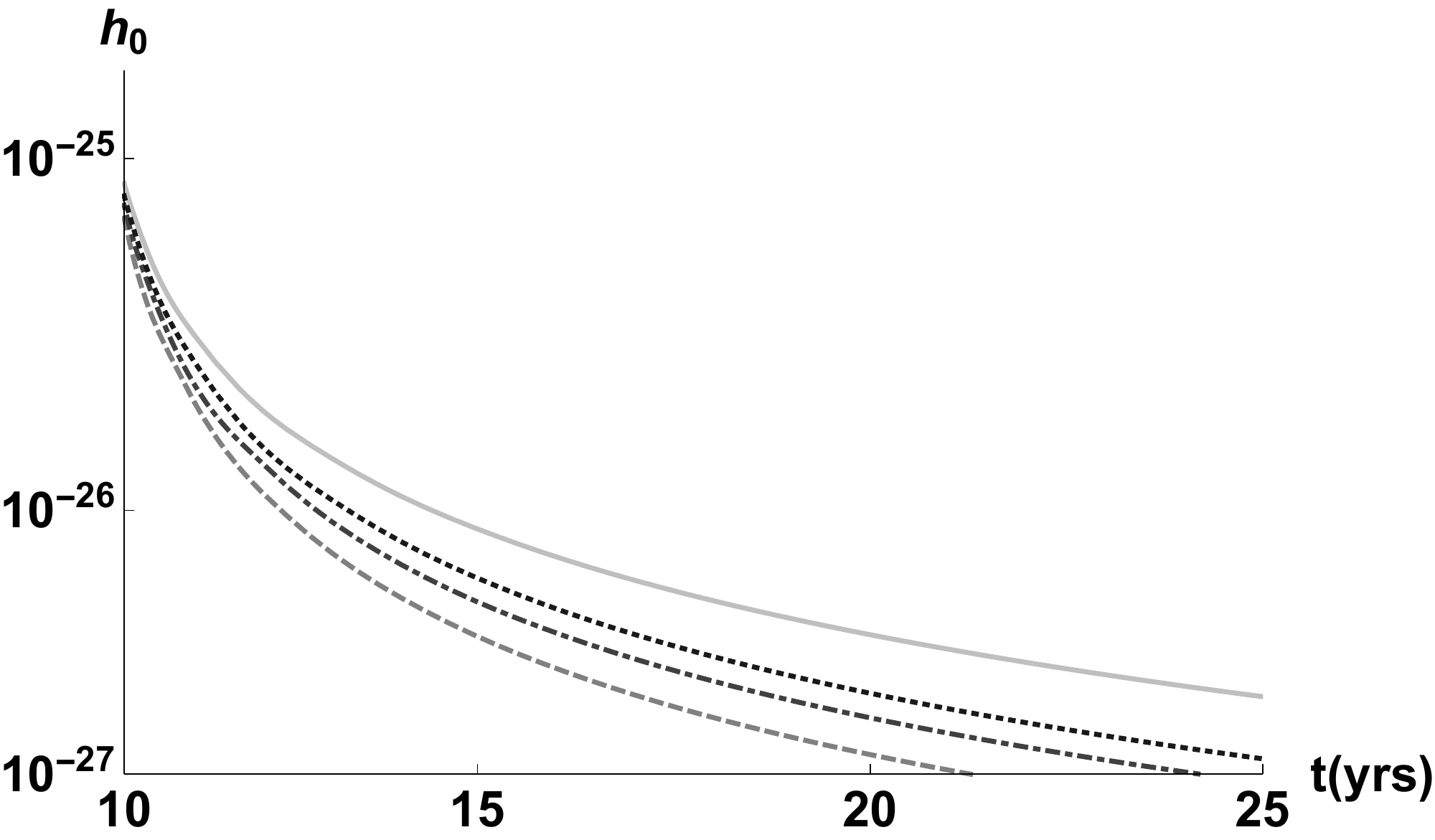}
    \caption{Evolution of the gravitational amplitude, a decade into the birth of the millisecond magnetar. The MMM charge has been fixed at $10^{-19}$, and the MMM masses have been taken at $15\,\mathrm{meV}$ (dashed), $20\,\mathrm{meV}$ (dot-dashed), and $25\,\mathrm{meV}$ (dotted). The evolution of the gravitational wave amplitude, when there are no MMMs is shown as a solid line. The initial conditions for the polar temperature ($4.5\times 10^{6}\,\mathrm{K}$), time period ($30\,\mathrm{ms}$) and mean magnetic field ($10^{16}\,\mathrm{G}$), were taken from representative values in the literature\,\cite{Vigano:2013lea,Metzger:2017wdz}. The distance to the source is assumed to be $1\,\mathrm{kpc}$. $\fD$ has been assumed to be $81$, corresponding to an $n=1$ polytropic equation of state. The amplitude must potentially be observable in third generation gravitational wave detectors, like the Einstein telescope, that is expected to have a sensitivity of $10^{-26}-10^{-27}$, in the $10-100\,\mathrm{Hz}$ frequency range, assuming integration times of one year\,\cite{Gualtieri:2010md, Hild:2010id,Glampedakis:2017nqy}.}
    \label{fig:gwevolve}
\end{figure}

Using Eq.\,(\ref{eq:hoamp}), the results of these numerical evolutions are displayed in Fig.\,\ref{fig:gwevolve}. As is clearly seen from these curves, the amplitudes deviate drastically from the conventional case, in the first few decades of the millisecond magnetar's birth. If $\tilde{\varepsilon}_{\text{\tiny{Q}}}$, or equivalently $\fD$, is even smaller, the main difference will be that the GW amplitudes will fall below their detectability much earlier in the epoch. As already mentioned, assuming a higher mean temperature would cause more conspicuous deviations with respect to conventional evolution. For the MMM masses and charges adopted in Fig.\,\ref{fig:gwevolve}, the neutron star temperature, for the time period displayed, is always higher than the respective critical temperatures $T_{\text{\tiny{C}}} \big(m,\xi,B(t)\big)$. Thus, for these parameter points, one does not expect, nor see, any relatively abrupt features in the gravitational wave amplitudes. Note also that the mean energetic arguments\,\cite{Hook:2017vyc} for these MMM masses, and corresponding limits on $\xi$ based on it, are still relevant. The thermal Schwinger pair-production rates are very prolific in the early epochs, but almost completely switch off once the magnetic field value decreases below the critical field value $\sim m^2/\xi g$; this happens after just a few decades. Thus, taken as an average over the entire lifetime of the magnetar, the mean energetic arguments should still furnish meaningful and interesting limits, while still being consistent with the enhanced rates and prominences in the early stages.

In general, as emphasised in subsection \ref{subsec:sppmmm}, one should expect to see comparatively abrupt features in the gravitational wave amplitude and frequency. They would have a distinct pattern, correlated with temperature and magnetic field evolution. The presence or absence of such abrupt patterns, in the GW waveform, would of course depend on the $(m,\xi)$ values of the MMMs that may exist in nature. More specifically, such abrupt patterns may appear if the mean temperature of the neutron star $T\ns(t)$ falls below the MMM critical temperature $T_C(t)$ at some point in time (equivalently, it may manifest through some evolution of a temperature gradient, across neutron star layers). After this cross-over there should be a relatively abrupt change in the MMM pair-production rates, and hence a relatively abrupt change in the gravitational wave amplitude and frequency evolution. Assume one is starting at an initial time $t_0$, with 
\begin{equation}
T\ns(t_0) > T_C(t_0)\; .
\end{equation}
For a cross-over to occur, a necessary criterion that the monotonically decreasing mean temperature and mean magnetic field profiles should satisfy, during some point subsequent to $t_0$, is 
\begin{equation}
 \frac{\dot{T}\ns(t)}{\dot{B}(t)}  \gtrsim \frac{\xi g }{2 m} \; .
\label{eq:abneccrit}
\end{equation}
Here, the dot denotes a first time derivative.
 
 For the gravitational waves to be detectable, such a crossing should also occur in the early stages of the millisecond magnetar's birth. Depending on the allowed values of $ \tilde{\varepsilon}_{\text{\tiny{Q}}}$, this may mean a time frame of seconds to decades, following birth. An MMM imprint detection is also more plausible during the early stages, since the internal magnetic fields are at their highest (implying large pair-production rates), and the temperatures are also varying rapidly (implying Eq.\,(\ref{eq:abneccrit}) is more prone to be satisfied). As seen from Fig.\,\ref{fig:CritTemp}, in the viable $\xi$ range, for MMM masses $m \lesssim 10^{-5}$, the critical temperatures can vary from $10^5-10^8\,\mathrm{K}$. As the neutron star is expected to cool from $10^{11}\,\mathrm{K}$ to $10^{6}\,\mathrm{K}$, over its initial phase of a few hundred years, if MMMs exist with the above mentioned masses and charges, they may leave imprints in the amplitude and frequency evolution that have a comparatively discontinuous character. During these epochs, they should also fall in the sensitivity ranges of future third generation gravitational wave detectors. 
 
If they exist, these MMM imprints on GWs, must be an almost universal feature across different newly-born millisecond magnetars. They must have a very unique pattern correlated with the temperature and magnetic field evolution, and hence should be potentially distinguishable from many other astrophysical phenomena. At the moment, it is difficult to quantitatively demonstrate this in a satisfactory manner, through an explicit rate computation and evolution, even in the simplified toy model. This is because, in the potentially interesting $(m,\xi)$  regions where such abrupt features may show up, we have $\xi g B/m^2 \gg 2 \pi$. Therefore, in these regions, all the known analytic expressions for thermal Schwinger pair production break down, and their applicability is unclear\,\cite{Gies:1998vt, Gies:1999vb,Ganguly:1998ys,Kim:2010qq, Brown:2015kgj, Medina,Gould:2017zwi,Gould,Gould:2018ovk,Korwar:2018euc,Draper:2018lyw}.
 
\section{Summary and conclusions}{\label{sec:summary}}
The search for continuous gravitational waves from neutron stars is well underway\,\cite{Abbott:2017ylp, Abbott:2017cvf, Authors:2019ztc}. Exotic particle states beyond the Standard Model have the potential to leave their imprints on these waveforms. In this work, we speculated on the effect of milli-magnetic monopoles on persistent gravitational wave signals, sourced by single neutron stars. 

Magnetic fields are known to cause distortions from spherical symmetry, in compact astrophysical objects, generating a quadrupole moment\,\cite{1953ApJ...118..116C,1954ApJ...119..407F}. If the magnetic and rotation axes are misaligned, this may produce detectable gravitational wave signals. Milli-magnetic monopoles may be copiously pair-produced in the extreme magnetic fields of neutron stars, such as magnetars; through the Schwinger pair-production mechanism\,\cite{Affleck:1981bma,Affleck:1981ag}. This causes an additional attenuation of the magnetic field, relative to conventional field decay mechanisms operational in a magnetar. Consequently, through a modification of the quadrupole moment time evolution, this may leave imprints in the continuous gravitational waves, during early stages of a neutron star's life. A time evolution of the neutron star quadrupole moment has been considered previously in other contexts\,\cite{Suvorov:2016hgr, deAraujo:2016wpz,deAraujo:2016ydk,deAraujo:2019xyn}. We found that the most promising candidate compact objects are a class of newly born magnetars, the so called millisecond magnetars\,\cite{Dai:1998bb,Zhang:2000wx,Rowlinson:2013ue,Giacomazzo:2013uua,2010ApJ...717..245K,2011MNRAS.413.2031M,Metzger:2017wdz}. In addition to deviations from conventional evolution, an imprint may potentially be present, as comparatively discontinuous features, in the gravitational waveform amplitude and frequency, in the early phases of a millisecond magnetar's life. Since the temperatures are rapidly evolving in the early stages, and the internal magnetic fields during these periods are also at their highest, these early times hold much promise. These signatures, if they exist as evidence for milli-magnetic monopoles, should be universally seen across new-born millisecond magnetars, with a very distinct pattern, and may therefore be potentially distinguishable from other astrophysical signatures.

A  more detailed implementation of the neutron star magneto-thermal evolution\,\cite{Aguilera:2007xk,Vigano:2013lea,2007A&A...470..303P}, incorporating milli-magnetic monopole non-perturbative production, should help further clarify and add to the ideas of the present study. Another crucial aspect is reaching a consensus on the functional form of the thermal Schwinger pair-production rates\,\cite{Brown:2015kgj, Medina,Gould:2017zwi,Gould,Gould:2018ovk,Korwar:2018euc,Draper:2018lyw} and striving to extend them to regions beyond the weak-field regime\,\cite{Affleck:1981bma,Kim:2000un, Kim:2003qp}. This would facilitate quantitative analyses in all regions of the viable $(m,\xi)$ parameter space, and directly probing the presence of abrupt features in the GW waveforms. Incorporating effects due to field inhomogeneities\,\cite{Dunne:2005sx} and finite chemical potentials\,\cite{Elmfors:1993wj, Zhang:2018hfd}, to account for the baryon environment and finite densities in a neutron star, would further sharpen future studies. Another crucial question is regarding how prevalent millisecond magnetars are\,\cite{2010ApJ...717..245K,2011MNRAS.413.2031M,Metzger:2017wdz}, and what their detection prospects are, across the lifetime of Advanced LIGO and future third generation GW detectors. We hope to address some of these in future works.
 
\begin{acknowledgments}
We thank Martin Hendry, Anson Hook, Adam Martin, Dipanjan Mitra, Sunil Mukhi and Prasad Subramanian for discussions. A.T. would like to thank the organisers of the Gordon Research Conference on Particle Physics 2019, where parts of this work were completed, and would also like to acknowledge support from an SERB Early Career Research Award.
\end{acknowledgments}


\bibliography{mmm_ellip_GW}
\bibliographystyle{JHEP}

\end{document}